\documentclass{entcs}

\usepackage{entcsmacro}
\usepackage{amssymb}
\usepackage{amsmath}
\usepackage[all]{xy}

\newtheorem{stp}{Step}
\newenvironment{step}{\vspace{-\lastskip}\par \addvspace{.6pc
    plus .2pc minus .1pc}\begin{stp}\rm}{\end{stp}\par\addvspace{.6pc
    plus .2pc minus .1pc}}

\newcommand{\ie}{\emph{i.e.}}
\newcommand{\eg}{\emph{e.g.}}

\newcommand{\Nat}{{\mathbb N}}
\newcommand{\Natp}{{\mathbb N_{>0}}}

\newcommand{\ximp}{\;\Rightarrow\;}
\newcommand{\xand}{\;\&\;}
\newcommand{\xor}{\text{ or }}
\newcommand{\xiff}{\text{ iff }}

\newcommand{\pnot}[1]{\lnot #1}
\newcommand{\pneg}[1]{\sim\!#1}
\newcommand{\pand}[2]{#1 \land #2}
\newcommand{\por}[2]{#1 \lor #2}
\newcommand{\pea}[2]{\langle#1\rangle #2}
\newcommand{\paa}[2]{[#1] #2}
\newcommand{\psp}[2]{#1;#2}
\newcommand{\pup}[2]{#1\cup#2}
\newcommand{\prp}[1]{#1*}
\newcommand{\pip}[1]{#1?}
\newcommand{\pimp}[2]{#1 \to #2}
\newcommand{\peq}[2]{#1 \leftrightarrow #2}

\newcommand{\pdl}{$PDL$}
\newcommand{\fml}{\mathrm{Fml}}
\newcommand{\prg}{\mathrm{Prg}}
\newcommand{\proptn}{\mathrm{AFml}}
\newcommand{\act}{\mathrm{APrg}}
\newcommand{\fea}{$\pea{}{}$}
\newcommand{\fean}{$\pea{\not\!\!a}{}$}
\newcommand{\fear}{$\pea{\prp{}}{}$}
\newcommand{\fmlea}{\mathrm{Fml}\pea{}{}}
\newcommand{\fmlean}{\mathrm{Fml}\pea{\not\!\!a}{}}
\newcommand{\fmler}{\mathrm{Fml}\pea{\prp{}}{}}

\newcommand{\evf}[1]{\tau_{#1}}
\newcommand{\evp}[1]{\rho_{#1}}
\newcommand{\psr}[2]{#1 \Vdash #2}

\newcommand{\prel}[3]{#1 \,#2\, #3}
\newcommand{\pnnf}[1]{\mathrm{nnf}(#1)}
\newcommand{\pcl}[1]{\mathrm{cl}(#1)}
\newcommand{\ppre}[1]{\mathrm{pre}(#1)}
\newcommand{\pzz}{\mathrel{\rightsquigarrow}}

\newcommand{\tnode}[3]{(#1 :: #2 :: #3)}
\newcommand{\thc}{\mathrm{HCr}}
\newcommand{\tnext}{\mathrm{Nx}}
\newcommand{\tbdia}{\mathrm{BD}}
\newcommand{\tbbox}{\mathrm{BB}}
\newcommand{\tmrk}{\mathrm{stat}}
\newcommand{\tuev}{\mathrm{uev}}

\newcommand{\talpha}{\boldsymbol{\alpha}}
\newcommand{\tbeta}{\boldsymbol{\beta}}
\newcommand{\trid}{$id$}
\newcommand{\trand}{$\pand{}{}$}
\newcommand{\tres}{$\pea{\psp{}{}}{}$}
\newcommand{\tras}{$\paa{\psp{}{}}{}$}
\newcommand{\trau}{$\paa{\pup{}{}}{}$}
\newcommand{\trar}{$\paa{\prp{}}{}$}
\newcommand{\trei}{$\pea{\pip{}}{}$}
\newcommand{\tror}{$\por{}{}$}
\newcommand{\treu}{$\pea{\pup{}{}}{}$}
\newcommand{\trero}{$\pea{\prp{}}{}_1$}
\newcommand{\trert}{$\pea{\prp{}}{}_2$}
\newcommand{\trai}{$\paa{\pip{}}{}$}
\newcommand{\trea}{$\langle\rangle$}

\newcommand{\ttrue}{\mathbf{unsat}}
\newcommand{\tloop}{\mathbf{barred}}
\newcommand{\tfalse}{\mathbf{open}}
\newcommand{\tmin}{\mathop{\mathrm{min}_{\bot}}}
\newcommand{\llen}{\mathop{\mathrm{len}}}
\newcommand{\tfean}{\mathop{\mathrm{tst}}}
\newcommand{\tbl}{\mathop{\mathrm{bl}}}

\newcommand{\ds}{\displaystyle\strut}
\newcommand{\ruleone}[3]{
\mbox{ {\bf #1} \ $\frac{\ds \mbox{$#2$}}{\ds \mbox{$#3$}}$}}

\newcommand{\tm}{T_{\mathrm{p}}}
\newcommand{\tl}{T_{\mathrm{l}}}
\newcommand{\fchn}{\sigma}

\newcommand{\extst}{\pea{\prp{(\pip{q})}}{(\pand{p}{\pnot{p}})}}
\newcommand{\exasp}{\paa{\prp{a}}{p}}
\newcommand{\exaasp}{\pea{\prp{(\psp{a}{a})}}{\pnot{p}}}

\def\lastname{Abate and Gor\'e and Widmann}

\begin{document}

\begin{frontmatter}
  \title{An On-the-fly Tableau-based Decision Procedure for \pdl{}-Satisfiability} 
  \author[Canberra]{Pietro Abate\thanksref{myemail}},
  \author[Canberra]{Rajeev Gor\'e\thanksref{coemail}} and
  \author[Nicta]{Florian Widmann\thanksref{cocoemail}\thanksref{NICTA}} 
  \address[Canberra]{Computer Sciences Laboratory
    \\The Australian National University
    \\ Canberra, Australia}
  \address[Nicta]{Computer Sciences Laboratory
    and Logic \& Computation Programme
    \\The Australian National University and NICTA
    \\ Canberra, Australia}
  \thanks[myemail]{Email:
    \href{mailto:Pietro.Abate@pps.jussieu.fr}
    {\texttt{\normalshape Pietro.Abate@pps.jussieu.fr}}} 
  \thanks[coemail]{Email:
    \href{mailto:Rajeev.Gore@rsise.anu.edu.au }
    {\texttt{\normalshape Rajeev.Gore@rsise.anu.edu.au }}}
  \thanks[cocoemail]{Email:
    \href{mailto:Florian.Widmann@rsise.anu.edu.au} 
    {\texttt{\normalshape Florian.Widmann@rsise.anu.edu.au}}}
  \thanks[NICTA]{National ICT Australia is
    funded by the Australian Government's Dept of Communications,
    Information Technology and the Arts and the Australian Research
    Council through Backing Australia's Ability and the ICT Centre of
    Excellence program.}
  \begin{abstract} 
    We present a tableau-based algorithm 
    for deciding satisfiability for propositional dynamic logic (\pdl{})
    which builds a finite rooted tree with ancestor loops
    and passes extra information from children to parents
    to separate good loops from bad loops during backtracking.
    It is easy to implement, with potential for parallelisation,
    because it constructs a pseudo-model ``on the fly''
    by exploring each tableau branch independently. But its
    worst-case behaviour is 2EXPTIME rather than EXPTIME.
    A prototype implementation in the TWB
    (\url{http://twb.rsise.anu.edu.au})
    is available.
  \end{abstract}
\end{frontmatter}

\section{Introduction}

Propositional dynamic logic (\pdl{}) is a logic
for reasoning about programs~\cite{pratt-semantical,fischer-ladner-dynamic}.
Its formulae consist of traditional Boolean formulae
plus ``action modalities'' built from a finite set of atomic programs
using sequential composition~$(\psp{}{})$, non-deterministic choice~$(\pup{}{})$,
repetition~$(\prp{})$, and test~$(\pip{})$.
The satisfiability problem for \pdl{}
is EXPTIME-complete~\cite{pratt-near-optimal-reasoning-about-action}.
Unlike EXPTIME-complete description logics
with algorithms exhibiting good average-case behaviour,
no decision procedures for \pdl{}-satisfiability are satisfactory
from both a theoretical (soundness and completeness)
and practical (average case behaviour) viewpoint as we explain below.

The earliest decision procedures for \pdl{}
are due to Fischer and Ladner~\cite{fischer-ladner-dynamic}
and Pratt~\cite{pratt-near-optimal-reasoning-about-action}.
Fischer and Ladner's method is impractical
because it first constructs 
the set of all consistent subsets of the set of all subformulae of the given formula,
which always requires exponential time in all cases.
On the other hand, Pratt~\cite{pratt-near-optimal-reasoning-about-action}
essentially builds a multi-pass (explained shortly) tableau method.
Most subsequent decision procedures for other fix-point logics
like propositional linear temporal logic (PLTL)~\cite{wolper-expressive},
computation tree logic (CTL)~\cite{ben-ari-pnueli-manna-branching,emerson-halpern-decision}
and the modal $\mu$-calculus~\cite{kozen-parikh}
trace back to Pratt~\cite{pratt-near-optimal-reasoning-about-action},
and they all share one main disadvantage as explained next.

In these multi-pass procedures,
a ``state'' is a node
which contains only diamond-like-formulae (``eventualities''),
box-like--formulae, atoms and negated atoms.
The first pass constructs a rooted tableau of nodes containing formula-sets,
but allows cross-branch arcs from a state~$n$ on one branch
to a (previously constructed) state~$m$ on a different branch
if applying the tableau construction to~$n$ would duplicate~$m$.
Thus the first pass constructs a ``pseudo-model''
which is a potentially exponential-sized cyclic graph
(rather than a cyclic tree where~$m$ would have to be an ancestor of~$n$).
The subsequent passes check that the ``pseudo-model'' is a real model
by pruning inconsistent nodes and pruning nodes containing ``unfulfilled eventualities''.

Although efficient model-checking techniques
can check the ``pseudo-model'' in time which is linear in its size,
these multi-pass methods can construct an exponential-sized cyclic graph needlessly.
One solution is to check for fulfilled eventualities ``on the fly'',
as the graph is built,
and although such methods exist for
model-checking~\cite{cleaveland,bhat-cleaveland-on-the-fly-model-checking},
we know of no such decision procedures for \pdl{}. 
The only implementation of a multiple-pass method for \pdl{}
that we know of is in LoTRec (\url{www.irit.fr/Lotrec})
but it is not optimal as it treats disjunctions naively.

Baader~\cite{baader-augmenting-transitive-closure}
gave a single-pass tableau-based decision procedure for a description logic
with role definitions involving union, composition and transitive closure of roles:
essentially \pdl{} without test.
His method constructs a (cyclic tree) tableau
using the semantics of the \pdl{} operators.
To separate ``good loops'' from ``bad loops'',
Baader must decide equality of regular languages,
a PSPACE-complete problem
which in practice may require exponential time.
Instead of solving these problems ``on the fly'',
they can be reduced to a simple check on the identity of states
in a deterministic minimal automaton
created from the positive regular expressions
appearing in the initial formula
during a pre-processing stage~\cite[page~27]{baader-augmenting-transitive-closure}.
But since the pre-computed automaton can be of exponential size,
this alternative may require exponential time needlessly. 
Baader's method is double-exponential in the worst-case.
The ``test'' construct is essential to express ``while'' loops
but creates a mutual recursion between the Boolean language
and the regular language.
It is not obvious to us how to extend Baader's method to ``test''.
DLP (\url{http://www.cs.bell-labs.com/cm/cs/who/pfps/dlp})
implements this method restricted to test-free formulae
where~$\prp{}$ applies only to atomic programs.

De Giacomo and Massacci~\cite{de-giacomo-massacci-tableaux-converse-pdl}
gave an optimal \pdl{}-satisfiability test
using labelled formulae like~$\sigma:\varphi$ to capture
that ``possible world~$\sigma$ makes formula~$\varphi$ true''.
They first give a NEXPTIME algorithm for deciding \pdl{}-satisfiability 
and then discuss ways to obtain an EXPTIME version using various known results.
But an actual EXPTIME algorithm,
and its soundness and completeness proofs,
are not given.
A deterministic implementation of their NEXPTIME algorithm by Schmidt and Tishkovsky
struck problems with nested stars,
but a solution is forthcoming~\cite{schmidt-tishkovsky-pdl-tableau}.

Other decision procedures for fix-point logics use resolution calculi,
translation methods, automata-theoretic methods, and game theoretic methods:
see~\cite{abate-gore-widmann-onepass-ctl-lpar} for references.
We know of no implementations for \pdl{} based on these methods.

Here, we give a sound, complete and terminating decision procedure for \pdl{}
with the following advantages and disadvantages:
\begin{description}
\item[\rm One-pass nature:] our method constructs a single-rooted finite tree
  (with loops from leaves to ancestors).
  As there are no cross-branch edges, 
  we can use depth-first, left-to-right search,
  reclaiming the space used for each branch via backtracking.
\item[\rm Proofs:] Full elementary proofs of soundness and completeness are available.
\item[\rm Ease of implementation:] our rules are easy to implement
  since our tableau nodes contain sets of formulae
  and some easily defined extra information
  whose manipulation requires only set intersection, set membership,
  and min/max on integers.
  But these low-level details make the rules cumbersome to describe.
\item[\rm Potential for optimisation:] there is potential
  to optimise our (tree) tableaux
  using successful techniques from (one-pass) tableaux
  for description logics~\cite{horrocks-patel-schneider-optimising-description-logic}.
\item[\rm Ease of generating counter-models:] the soundness proof immediately gives
  an effective procedure for turning an ``open'' tableau into a \pdl{}-model.
\item[\rm Ease of generating proofs:] unlike existing Gentzen calculi
  for fix-point logics~\cite{alberucci-jaeger-lck,kretz-studer-jaeger-lck-cut-free},
  our tableau calculus gives a cut-free Gentzen-style calculus
  with ``cyclic proofs'' with an optimal rather than worst-case bound
  for the finitised omega rule.
\item[\rm Potential for parallelisation:] our rules build the branches independently
  but combine their results during backtracking,
  enabling a parallel implementation.
\item[\rm Prototype:] a (sequential) prototype implementation
  in the Tableau Work Bench
  (\url{twb.rsise.anu.edu.au})
  allows to test arbitrary \pdl{} formulae over the web.
\item[\rm Complexity:] our method has worst-case double-exponential time complexity.
\item[\rm Generality:] Our method for \pdl{}
  fits into a class of similar ``one pass'' methods
  for other fix-point logics like
  PLTL~\cite{schwendimann-one-pass} and
  CTL~\cite{abate-gore-widmann-onepass-ctl-lpar}. 
  Further experimental work is required to determine
  if our methods can be optimised to exhibit good average-case behaviour
  using techniques like sound global caching~\cite{gore-nguyen-exptime-alc}.
\end{description}

\section{Syntax, Semantics and Hintikka Structures}

\begin{definition}
  Let~$\proptn$ and~$\act$ be two disjoint and countably infinite sets of propositional atoms
  and \emph{atomic programs}, respectively.
  The set~$\fml$ of all formulae and the set~$\prg$ of all \emph{programs}
  are defined inductively as follows:
  \begin{enumerate}
  \item $\proptn \subseteq \fml$ and~$\act \subseteq \prg$
  \item if~$\varphi, \psi \in \fml$ then~$\pnot{\varphi} \in \fml$
    and~$\pand{\varphi}{\psi} \in \fml$ and~$\por{\varphi}{\psi} \in \fml$ and~$\pip{\varphi} \in \prg$
  \item if~$\varphi \in \fml$ and~$\alpha \in \prg$ then~$\pea{\alpha}{\varphi}\in \fml$ and~$\paa{\alpha}{\varphi} \in \fml$
  \item if~$\alpha \in \prg$ and $\beta \in \prg$ then~$(\psp{\alpha}{\beta}) \in \prg$ and~$\pup{\alpha}{\beta} \in \prg$ and~$\prp{\alpha} \in \prg$.
  \end{enumerate}
  Let~$p, q$ range over members of~$\proptn$
  and~$a, b$ range over members of~$\act$.
  A \fea{}-formula is any formula~$\pea{\alpha}{\varphi}$,
  a \fean{}-formula is a \fea{}-formula 
  $\pea{\alpha}{\varphi}$ with~$\alpha \notin \act$,
  and a \fear{}-formula is any formula~$\pea{\prp{\alpha}}{\varphi}$.
  $\fmlea$ is the set of all \fea{}-formulae,
  $\fmlean$ is the set of all \fean{}-formula,
  and~$\fmler$ is the set of all \fear{}-formulae.
\end{definition}

\begin{definition}
  A \emph{transition frame} is a pair~$(W,R)$
  where~$W$ is a non-empty set of worlds and~$R$ a function
  that maps each atomic program~$a$ to a binary relation~$R_a$ over~$W$.
  A \emph{model}~$(W,R,V)$ is a transition frame~$(W,R)$
  and a valuation function~$V: \proptn \to 2^W$
  mapping each atomic proposition~$p$ to
  a set~$V(p)$ of worlds.
\end{definition}

\begin{definition}
  Let~$M = (W,R,V)$ be a model.
  The functions~$\evf{M}: \fml \to 2^W$ and~$\evp{M}: \prg \to 2^{W \times W}$
  are defined inductively as follows:
  \begin{displaymath}
    \begin{array}{l@{\extracolsep{1cm}}r}
      \multicolumn{2}{c}{
        \evf{M}(p)
        :=
        V(p)
        \hfill
        \evp{M}(a)
        :=
        R_a
        \hfill
        \evf{M}(\pnot{\varphi})
        := 
        W \setminus \evf{M}(\varphi)}\\
      \evf{M}(\pand{\varphi}{\psi})
      := 
      \evf{M}(\varphi) \cap \evf{M}(\psi)
      &
      \evf{M}(\por{\varphi}{\psi})
      := \evf{M}(\varphi) \cup \evf{M}(\psi)\\
      \multicolumn{2}{c}{
        \evf{M}(\paa{\alpha}{\varphi})
        :=
        \{ w \mid \forall v \in W.\: (w, v) \in \evp{M}(\alpha) \ximp v \in \evf{M}(\varphi) \}}\\
      \multicolumn{2}{c}{
        \evf{M}(\pea{\alpha}{\varphi})
        := 
        \{ w \mid \exists v \in W.\: (w, v) \in \evp{M}(\alpha) \xand v \in \evf{M}(\varphi) \}}\\
      \evp{M}(\pup{\alpha}{\beta})
      := 
      \evp{M}(\alpha) \cup \evp{M}(\beta)
      &
      \evp{M}(\pip{\varphi})
      := \{ (w, w) \mid w \in \evf{M}(\varphi) \}\\
      \multicolumn{2}{c}{
        \evp{M}(\psp{\alpha}{\beta})
        := 
        \{ (w, v) \mid \exists u \in W.\: (w, u) \in \evp{M}(\alpha) \xand (u, v) \in \evp{M}(\beta) \}}\\
      \multicolumn{2}{c}{
        \evp{M}(\prp{\alpha})
        :=
        \big\{ (w, v) \mid
        \begin{array}[t]{l}
          \exists k \in \Nat.
          \exists w_0, \dotsc, w_k \in W.\: \big( w_0 = w \xand w_k = v \xand\\
          \forall i \in \{ 0, \dotsc, k-1 \}.\: (w_i, w_{i+1}) \in \evp{M}(\alpha) \big) \big\}
        \end{array}}\\
    \end{array}
  \end{displaymath}
  For~$w \in W$ and~$\varphi \in \fml$, we write
  $\psr{M,w}{\varphi} \xiff w \in \evf{M}(\varphi)$.
\end{definition}

\begin{definition}
  Formula~$\varphi \in \fml$ is \emph{satisfiable}
  iff there is a model~$M = (W,R,V)$ and a~$w \in W$
  such that~$\psr{M,w}{\varphi}$.
  Formula~$\varphi \in \fml$ is \emph{valid} iff~$\pnot{\varphi}$ is not satisfiable.
\end{definition}

\begin{definition}
  Formula~$\varphi \in \fml$ is in \emph{negation normal form}
  if~$\pnot{}$ appears only immediately before propositional atoms.
  For every~$\varphi \in \fml$,
  we obtain a formula~$\pnnf{\varphi}$ in negation normal form
  by pushing negations inward repeatedly
  (\eg{} using de Morgan's laws)
  so~$\peq{\varphi}{\pnnf{\varphi}}$ is valid.
  We define~$\pneg{\varphi} := \pnnf{\pnot{\varphi}}$.
\end{definition}

\begin{table}[t]
  \caption{Smullyan's $\talpha$- and $\tbeta$-notation to classify formulae}
  \label{tab_alphabeta}
  \begin{center}
    \begin{tabular}{|c|c|c|c|c|c|c|}
      \hline
      $\talpha$ 
      & $\pand{\varphi}{\psi}$ 
      & $\paa{\pup{\alpha}{\beta}}{\varphi}$ 
      & $\paa{\prp{\alpha}}{\varphi}$ 
      & $\pea{\pip{\psi}}{\varphi}$ 
      & $\pea{\psp{\alpha}{\beta}}{\varphi}$ 
      & $\paa{\psp{\alpha}{\beta}}{\varphi}$ 
      \\ \hline
      $\talpha_1$ 
      & $\varphi$ 
      & $\paa{\alpha}{\varphi}$ 
      & $\varphi$ 
      & $\varphi$ 
      & $\pea{\alpha}{\pea{\beta}{\varphi}}$ 
      & $\paa{\alpha}{\paa{\beta}{\varphi}}$ 
      \\ \hline
      $\talpha_2$
      & $\psi$
      & $\paa{\beta}{\varphi}$
      & $\paa{\alpha}{\paa{\prp{\alpha}}{\varphi}}$
      & $\psi$
      & 
      & 
      \\ \hline
      \end{tabular}
      \begin{tabular}{|c|c|c|c|c|}
      \hline
      $\tbeta$ 
      & $\por{\varphi}{\psi}$ 
      & $\pea{\pup{\alpha}{\beta}}{\varphi}$ 
      & $\pea{\prp{\alpha}}{\varphi}$ 
      & $\paa{\pip{\psi}}{\varphi}$ 
      \\ \hline
      $\tbeta_1$ 
      & $\varphi$ 
      & $\pea{\alpha}{\varphi}$ 
      & $\varphi$ 
      & $\varphi$ 
      \\ \hline
      $\tbeta_2$
      & $\psi$
      & $\pea{\beta}{\varphi}$
      & $\pea{\alpha}{\pea{\prp{\alpha}}{\varphi}}$
      & $\pneg{\psi}$
      \\ \hline
    \end{tabular}
  \end{center}
\end{table}
We use Smullyan's $\talpha$/$\tbeta$-notation to categorise formulae
via Table~\ref{tab_alphabeta}
and use bolding to differentiate it from the use of~$\alpha$ and~$\beta$ as members of~$\prg$.
So if~$\talpha$~(respectively $\tbeta$) is any formula pattern in the first row
then~$\talpha_1$ and~$\talpha_2$ (respectively~$\tbeta_1$ and~$\tbeta_2$)
are its corresponding patterns in the second and third row.
\begin{proposition}
  \label{prop_axioms}
  All formulae~$\peq{\talpha}{\pand{\talpha_1}{\talpha_2}}$
  and~$\peq{\tbeta}{\por{\tbeta_1}{\tbeta_2}}$ in
  Table~\ref{tab_alphabeta} are valid.
\end{proposition}

\begin{definition}
  A \emph{structure}~$(W,R,L)$ $[$for~$\varphi \in \fml]$
  is a transition frame~$(W,R)$
  and a labelling function~$L: W \to 2^{\fml}$ which associates with each world~$w \in W$
  a set~$L(w)$ of formulae $[$and has~$\varphi \in L(v)$ for some world~$v \in W]$.
\end{definition}

\begin{definition}
  For a given~$\varphi \in \fml$
  the (infinite) set~$\ppre{\varphi}$ is defined as:
  \[ \ppre{\varphi} := \{ \psi \in \fml \mid \exists k \in \Nat.\: \exists \alpha_1, \dotsc, \alpha_k \in \prg.\: \psi = \pea{\alpha_1}{\dotsc \pea{\alpha_k}{\varphi}} \} \enspace.\]
  For all formulae~$\varphi$ and~$\psi$,
  the binary relation~$\pzz$ on formulae is defined as:
  $\varphi \pzz \psi$ iff (exactly) one of the following conditions is true:
  \begin{itemize}
  \item $\exists \chi \in \fml . \exists \alpha, \beta \in \prg.\: \varphi = \pea{\psp{\alpha}{\beta}}{\chi} \xand \psi = \pea{\alpha}{\pea{\beta}{\chi}}$
  \item $\exists \chi \in \fml . \exists \alpha, \beta \in \prg.\: \varphi = \pea{\pup{\alpha}{\beta}}{\chi}
    \xand \big( \psi = \pea{\alpha}{\chi} \xor \psi = \pea{\beta}{\chi} \big)$
  \item $\exists \chi \in \fml . \exists \alpha \in \prg.\:
    \varphi = \pea{\prp{\alpha}}{\chi} \xand
    \big( \psi = \chi \xor \psi = \pea{\alpha}{\pea{\prp{\alpha}}{\chi}} \big)$
  \item $\exists \chi, \phi \in \fml.\: \varphi = \pea{\pip{\phi}}{\chi} \xand \psi = \chi \enspace.$
  \end{itemize}
\end{definition}
Intuitively, using Table~\ref{tab_alphabeta},
the~``$\pzz$'' relates a \fean{}-formulae~$\talpha$ (respectively~$\tbeta$),
to~$\talpha_1$ (respectively~$\tbeta_1$ and~$\tbeta_2$)
while~$\ppre{\varphi}$ captures
that~$\pea{\prp{\alpha}}{\varphi}$ can be ``reduced'' to~$\pea{\alpha}{\pea{\prp{\alpha}}{\varphi}}$,
which can be reduced to~$\pea{\alpha_1}{\dotsc \pea{\alpha_k}{\pea{\prp{\alpha}}{\varphi}}}$.
Note that~$\varphi \in \ppre{\varphi}$.

\begin{definition}
  Let~$H = (W,R,L)$ be a structure, $\varphi \in \fml$ a formula, $\beta \in \prg$ a program, and~$w \in W$ a state.
  A \emph{fulfilling chain for~$(\varphi, \beta, w)$ in~$H$}
  is a finite sequence~$(w_0, \psi_0), \dotsc, (w_n, \psi_n)$ of world-formula pairs with~$n \geq 0$ such that:
  \begin{itemize}
  \item $w_i \in W$, $\psi_i \in \ppre{\varphi}$, and~$\psi_i \in L(w_i)$ for all~$0 \leq i \leq n$
  \item $w_0 = w$, $\psi_0 = \pea{\beta}{\varphi}$, $\psi_n = \varphi$, and~$\psi_i \neq \varphi$ for all~$0 \leq i \leq n-1$
  \item for all~$0 \leq i \leq n-1$,
    if~$\psi_i = \pea{a}{\chi}$ for some~$a \in \act$ and~$\chi \in \fml$
    then~$\psi_{i+1} = \chi$ and~$\prel{w_i}{R_a}{w_{i+1}}$;
    otherwise~$\psi_i \pzz \psi_{i+1}$ and~$w_i = w_{i+1}$.
  \end{itemize}
\end{definition}
Each~$\psi_i$ is in~$L(w_i)$,
the chain starts at~$(w_0, \pea{\beta}{\varphi})$, ends at~$(w_n, \varphi)$,
and no other~$w_i$ is paired with~$\varphi$.
Formulae~$\psi_i, \psi_{i+1}$ are~$\pzz$-related 
and corresponding worlds~$w_i, w_{i+1}$ are equal
unless~$\psi_i = \pea{a}{\chi}$,
in which case~$\psi_{i+1} = \chi$ and~$\prel{w_i}{R_a}{w_{i+1}}$.
Thus eventuality~$\pea{\beta}{\varphi} \in w_0$ is fulfilled by~$\varphi \in w_n$
and~$w_n$ is $\beta$-reachable from~$w_0$.

\begin{definition}
  A \emph{pre-Hintikka structure}~$H = (W,R,L)$ $[$for~$\varphi \in \fml]$
  is a structure $[$for~$\varphi]$
  that satisfies~H1-H5 (below) for every~$w \in W$
  where~$\talpha$ and~$\tbeta$ are formulae as defined in Table~\ref{tab_alphabeta}.
  A \emph{Hintikka structure}~$H = (W,R,L)$ $[$for~$\varphi \in \fml]$
  is a pre-Hintikka structure $[$for~$\varphi]$
  that additionally satisfies~H6 below:
  \begin{displaymath}
    \begin{array}{ll}
      \mathrm{H1:} & \pnot{p} \in L(w) \ximp p \not\in L(w)\\
      \mathrm{H2:} & \talpha \in L(w) \ximp \talpha_1 \in L(w) \xand \talpha_2 \in L(w)\\
      \mathrm{H3:} & \tbeta \in L(w) \ximp \tbeta_1 \in L(w) \xor \tbeta_2 \in L(w)\\
      \mathrm{H4:} & \pea{a}{\varphi} \in L(w) \ximp \exists v \in W.\: \prel{w}{R_a}{v} \xand \varphi \in L(v)\\
      \mathrm{H5:} & \paa{a}{\varphi} \in L(w) \ximp \forall v \in W.\: \prel{w}{R_a}{v} \ximp \varphi \in L(v)\\
      \mathrm{H6:} & \pea{\prp{\alpha}}{\varphi} \in L(w) \ximp
      \text{there exists a fulfilling chain for~$(\varphi, \prp{\alpha}, w)$ in~$H$} \enspace.
    \end{array}
  \end{displaymath}
\end{definition}
H3 ``locally unwinds'' the fix-point semantics of~$\pea{\prp{\alpha}}{\varphi}$,
but does not guarantee a \emph{least} fix-point which 
requires~$\varphi$ be true eventually.
H6 ``globally''
ensures all \fear{}-formulae are fulfilled.
H2 captures the \emph{greatest} fix-point 
semantics of~$\paa{\prp{\alpha}}{\varphi}$.

\begin{theorem}
  \label{theo_satisfiable}
  A formula~$\varphi \in \fml$ in negation normal form is satisfiable
  iff there exists a Hintikka structure for~$\varphi$.
\end{theorem}

\section{An Overview of the Algorithm}
\label{sec_overview}

To track unfulfilled eventualities and to avoid ``at a world'' cycles,
our algorithm stores additional information in each tableau node
using \emph{histories} and
\emph{variables}~\cite{schwendimann-one-pass}.
Histories are passed from parents to children and
variables from children to parents.

Our algorithm starts at a root
containing a given formula~$\phi$
and some default history values.
It builds a tree by repeatedly applying $\talpha$-/$\tbeta$-rules
to decompose formulae via the semantics of \pdl{}.
The $\tbeta$-rule for~$\pea{\prp{\alpha}}{\varphi}$ has a left child
that fulfils this eventuality by reducing it to~$\varphi$,
and a right child
that procrastinates fulfilment by ``reducing'' it to~$\pea{\alpha}{\pea{\prp{\alpha}}{\varphi}}$.
The rules modify the histories and variables as appropriate
for their intended purpose.

But naive application of the $\talpha$-/$\tbeta$-rules to
formulae like~$\pea{\prp{\prp{a}}}{\varphi}$
with nested stars can lead to ``at a world'' cycles: e.g.\
$
  \pea{\prp{\prp{a}}}{\varphi},
  \cdots ,
  \pea{\prp{a}}{\pea{\prp{\prp{a}}}{\varphi}},
  \cdots ,
  \pea{\prp{\prp{a}}}{\varphi}
$.
A solution is to use the histories to reduce one particular~$\pea{\alpha}{}$-formula
until~$\alpha$ becomes atomic by forcing the rules to concentrate on
this task, and 
to block previously reduced diamonds and boxes
if they lead to ``at a world'' cycles.
The application of $\talpha$/$\tbeta$-rules stops
when all non-blocked leaves contain only atoms, negated atoms,
and all \fea{}-formulae and all $\paa{}{}$-formulae
begin with outermost atomic programs only.

For each such leaf node~$l$, and for each $\pea{a}{\xi}$-formula in~$l$,
the \trea{}-rule creates a successor node containing~$\{ \xi \} \cup \Delta$,
where~$\Delta = \{\psi \mid \paa{a}{\psi} \in l\}$.
These successors are then saturated to produce new leaves
using the $\talpha$- and $\tbeta$-rules,
and the \trea{}-rule creates the successors of these new leaves, and so on.

If left unchecked, this procedure can produce infinite branches
since the same successors can be created again and again on the same branch.
To obtain termination,
the \trea{}-rule creates a successor containing~$\{ \xi \} \cup \Delta$ for $l$
only if 
this successor has not already been created previously higher up on the current branch.

So if the successor~$\{ \xi \} \cup \Delta$ exists already,
the current branch is ``blocked'' from re-creating it.
The resulting loop may be ``bad'' since every $\tbeta$-node
on this branch for an eventuality~$\pea{\prp{\alpha}}{\varphi}$ may procrastinate,
so~$\pea{\prp{\alpha}}{\varphi}$ is never fulfilled.
To track this potentially unfulfilled eventuality,
we assign the height of the blocking node to the pair~$(\xi, \pea{\prp{\alpha}}{\varphi})$ 
via a variable~$\tuev$ as long as~$\xi$ is a decomposition of~$\pea{\prp{\alpha}}{\varphi}$.

During backtracking, our rules ``merge'' the~$\tuev$ entries of the children
and also modify the resulting~$\tuev$
to reverse-track the decomposition of~$\pea{\prp{\alpha}}{\varphi}$.
In particular, a~$\tuev$ entry becomes undefined at a node
if the eventuality it tracks can be fulfilled in the sub-tableau rooted at this node.
Conversely, if a node at height~$h$ receives a~$\tuev$ entry with value at least~$h$
then the eventuality tracked by this~$\tuev$ entry definitely cannot be fulfilled,
so the parent of this (blocking) node is then unsatisfiable.

Whether or not the initial formula~$\phi$ is satisfiable
is determined by the status of the root node.
Due to technicalities caused by ``at a world'' cycles,
the status can be one of the values ``unsatisfiable'', ``open'' or ``barred'' (to be explained later).
The initial formula~$\phi$ is \pdl{}-satisfiable iff the status of the root node is ``open''.

\section{A One-pass Tableau Algorithm for \pdl{}}

\begin{definition}
  A \emph{tableau node}~$x$ is of the form $\tnode{\Gamma}{\thc, \tnext, \tbdia, \tbbox}{\tmrk, \tuev}$ where:
  $\Gamma$ is a set of formulae;
  $\thc$ is a list of pairs~$(\varphi, \Delta)$ where~$\Delta$ is a set of formulae and~$\varphi \in \Delta$;
  $\tnext$ is either~$\bot$ or a formula designated to be the principal formula of the rule applied to~$x$;
  $\tbdia$ is the set of ``\underline{B}locked \underline{D}iamonds'';
  $\tbbox$ is the set of ``\underline{B}locked \underline{B}oxes'';
  $\tmrk$ has one of the values~$\ttrue$, $\tfalse$, or~$\tloop$; and
  $\tuev$ is a partial function from $\fmlea \!\times \fmler$
  to~$\Natp$ (the positive natural numbers).
\end{definition}

\begin{definition}
  A \emph{tableau} for a formula set~$\Gamma \subseteq \fml$ and histories~$\thc$, $\tnext$, $\tbdia$, and~$\tbbox$
  is a tree of tableau nodes with root $\tnode{\Gamma}{\thc, \tnext, \tbdia, \tbbox}{\tmrk, \tuev}$
  where the children of a node~$x$ are obtained by a single application of a rule to~$x$
  (\ie{} only one rule can be applied to a node)
  but where the parent can inherit some information from the children.
  A tableau is \emph{expanded} if no rules can be applied to any of its leaves.
  On any branch of a tableau, a node~$t$ is an \emph{ancestor} of a node~$s$
  iff~$t$ lies above~$s$ on the unique path from the root down to~$s$.
\end{definition}

The list~$\thc$ is a history for detecting ancestor-loops and guarantees termination. 
The choice of principal formula is free if $\tnext = \bot$, but is pre-determined as the
formula in~$\tnext$ otherwise.
When a diamond formula in the parent is decomposed
to give a formula~$\varphi \in \fmlean$ in the current node,
we set the~$\tnext$-value of the child to $\varphi$ to  ensure that~$\varphi$ is decomposed next.
Together with the histories~$\tbdia$ and~$\tbbox$,
this allows us to block~$\pea{\prp{\alpha}}{}$-formulae and~$\paa{\prp{\alpha}}{}$-formulae
from creating ``at a world'' cycles.
The variables~$\tmrk$ and~$\tuev$ have their values
determined by the children of a node.
Formally, $\tmrk = \ttrue$ at node~$x$
if~$x$ is definitely unsatisfiable.
Informally, $\tmrk = \tloop$ if all descendants of node~$x$ are unsatisfiable
or lead to an ``at a world'' cycle.
Finally, $\tmrk = \tfalse$ indicates that the node is potentially satisfiable,
but as it may be on a loop,
this is something which we can determine only later as we backtrack towards the root. 

\begin{definition}
  The partial function~$\tuev_{\bot} : \fmlea \times \fmler \rightharpoonup \Natp$ is the constant function
  that is undefined for all pairs of formulae: \ie{}~$\forall \psi_1, \psi_2.\: \tuev_{\bot}(\psi_1, \psi_2) = \bot$.
  The partial functions~$\tfean : \fml \rightharpoonup \fml$ and~$\tbl : \fml \times 2^{\fml} \rightharpoonup 2^{\fml}$ are defined as:
  \begin{displaymath}
    \begin{array}{l@{\extracolsep{1cm}}r}
      \tfean(\chi) := 
      \left\{
        \begin{array}{ll}
          \chi & \text{ if } \chi \in \fmlean\\
          \bot & \text{ otherwise}
        \end{array}
      \right .
      &
      \tbl(\chi, \Gamma) := 
      \left\{
        \begin{array}{ll}
          \Gamma & \text{ if } \chi \in \fmlean\\
          \emptyset & \text{ otherwise.}
        \end{array}
      \right .
    \end{array}
  \end{displaymath}
\end{definition}
The function~$\tfean{}$ returns~$\bot$
when the formula being tested is not a \fea{}-formula,
or is a \fea{}-formula but its program is atomic.
The function~$\tuev$ tracks
\underline{u}nfulfilled \underline{ev}entualities,
so~$\tuev_{\bot}$ flags that all eventualities are fulfilled,
and~$\tuev(\chi_1, \chi_2)$ defined flags a potentially unfulfilled eventuality.
If a node has~$\tmrk = \ttrue$ or~$\tmrk = \tloop$
then its~$\tuev$ is irrelevant so it is arbitrarily set to~$\tuev_{\bot}$.

\subsection{The Rules}

We use~$\Gamma$ and~$\Delta$ for sets of formulae
and write~$\varphi_1 \ , \ \dotsc \ , \ \varphi_n \ , \ \Delta_1 \ , \ \dotsc \ , \ \Delta_m$ for the
partition~$\{ \varphi_1 \} \uplus \dotsb \uplus \{ \varphi_n \} \uplus \Delta_1 \uplus \dotsb \uplus \Delta_m$ of formulae in a node.
To save space, we often omit histories/variables which are passed
unchanged from parents/children to children/parents.
Most rules are applicable only if some side-conditions hold,
and most involve actions that change histories downwards or variables upwards.

\paragraph{Terminal Rules.}

\begin{center}
  \ruleone{(\trid{})}
  {\tnode{\Gamma}{\dotsb}{\tmrk, \tuev}}{}
  $\quad \{ p, \pnot{p} \} \subseteq \Gamma \text{ for some } p \in \proptn$
\end{center}
Action for~(\trid{}): $\tmrk := \ttrue$ and~$\tuev := \tuev_{\bot}$.
\begin{center}
  \ruleone{(\trert{})}
  {\tnode{\pea{\prp{\alpha}}{\varphi} , \; \Gamma}{\tnext, \tbdia}{\tmrk, \tuev}}{}
  $\quad \tnext \in \{ \bot, \pea{\prp{\alpha}}{\varphi} \} \xand \pea{\prp{\alpha}}{\varphi} \in \tbdia$
\end{center}
Action for~(\trert{}): $\tmrk := \tloop$ and~$\tuev := \tuev_{\bot}$.

An \trid{}-node is clearly unsatisfiable.
The principal formula of the \trert{}-rule is unfulfillable
because it causes an ``at a world'' cycle,
so this rule terminates the current branch. Note both rules may be
applicable to a node.

\paragraph{Linear ($\talpha$) Rules.}

\begin{tabular}[c]{l@{\extracolsep{1.5cm}}l}
  \ruleone{(\trand{})}
  {\tnode{\pand{\varphi}{\psi} , \; \Gamma}{\tnext}{\tuev}}
  {\tnode{\varphi , \; \psi , \; \Gamma}{\tnext}{\tuev_1}}
  &
  \ruleone{(\trau{})}
  {\tnode{\paa{\pup{\alpha}{\beta}}{\varphi} , \; \Gamma}{\tnext}{\tuev}}
  {\tnode{\paa{\alpha}{\varphi} , \; \paa{\beta}{\varphi} , \; \Gamma}{\tnext}{\tuev_1}}
  \\[2em]
  \ruleone{(\tras{})}
  {\tnode{\paa{\psp{\alpha}{\beta}}{\varphi} , \; \Gamma}{\tnext}{\tuev}}
  {\tnode{\paa{\alpha}{\paa{\beta}{\varphi}} , \; \Gamma}{\tnext}{\tuev_1}}
  &
  \ruleone{(\trar{})}
  {\tnode{\paa{\prp{\alpha}}{\varphi} , \; \Gamma}{\tnext, \tbbox}{\tuev}}
  {\tnode{\Gamma_1}{\tnext, \tbbox_1}{\tuev_1}}
  \\[1em]
\end{tabular}
\begin{flushleft}
  Common Side Condition: $\quad \tnext = \bot$.
  
  Common Action:
  $\tuev(\chi_1, \chi_2) := \tuev_1(\chi_1, \chi_2)$
  if $\chi_1 \in \Gamma$ 
  else $\tuev(\chi_1, \chi_2) := \bot$.
  
  Extra Action for~(\trar{}):
  \begin{array}[t]{l}
    \Gamma_1 := \Gamma
    \text{ if } \paa{\prp{\alpha}}{\varphi} \in \tbbox
    \text{ else } \Gamma_1 := \{\varphi\} \cup \{\paa{\alpha}{\paa{\prp{\alpha}}{\varphi}}\} \cup \Gamma,\\
    \tbbox_1 := \big\{ \paa{\prp{\alpha}}{\varphi} \big\} \cup \tbbox.
  \end{array}
\end{flushleft}

Most rules are standard but for the histories
since they just capture the transformations in Table~\ref{tab_alphabeta}.
The \trar{}-rule just deletes~$\paa{\prp{\alpha}}{\varphi}$ 
if~$\paa{\prp{\alpha}}{\varphi} \in \tbbox$ since this indicates
that it has already been expanded once ``at this world''.
Otherwise it captures the fix-point nature of~$\paa{\prp{\alpha}}{\varphi}$ via Prop.~\ref{prop_axioms}
and then puts~$\paa{\prp{\alpha}}{\varphi}$ into~$\tbbox_1$.

The next two rules have individual side-conditions and actions as shown.
\begin{center}
  \ruleone{(\tres{})}
  {\tnode{\pea{\psp{\alpha}{\beta}}{\varphi} , \; \Gamma}{\tnext, \tbdia}{\tuev}}
  {\tnode{\pea{\alpha}{\pea{\beta}{\varphi}} , \; \Gamma}{\tnext_1, \tbdia_1}{\tuev_1}}
  $\quad\tnext \in \{ \bot, \pea{\psp{\alpha}{\beta}}{\varphi} \}$
\end{center}
\begin{flushleft}
  Actions for~(\tres{}): \ \\
  \begin{tabular}[c]{lr}
    \begin{minipage}[l]{0.28\linewidth}
      \begin{eqnarray*}
        \tnext_1 & := & \tfean\big( \pea{\alpha}{\pea{\beta}{\varphi}} \big)
        \\ \\
        \tbdia_1 & := & \tbl\big( \pea{\alpha}{\pea{\beta}{\varphi}}, \tbdia \big)
      \end{eqnarray*}
    \end{minipage}
    &
    \begin{minipage}[l]{0.72\linewidth}
      \begin{eqnarray*}
        \tuev(\chi_1, \chi_2) & := &
        \left\{
          \begin{array}{ll}
            \tuev_1(\pea{\alpha}{\pea{\beta}{\varphi}}, \chi_2) & \text{ if } \chi_1 = \pea{\psp{\alpha}{\beta}}{\varphi}\\
            \tuev_1(\chi_1, \chi_2) & \text{ if } \chi_1 \in \Gamma\\
            \bot & \text{ otherwise}
          \end{array}
        \right .
      \end{eqnarray*}
    \end{minipage}
  \end{tabular}
\end{flushleft}

\begin{center}
  \ruleone{(\trei{})}
  {\tnode{\pea{\pip{\psi}}{\varphi} , \; \Gamma}{\tnext, \tbdia_1}{\tuev}}
  {\tnode{\psi , \; \varphi , \; \Gamma}{\tnext_1, \tbdia_1}{\tuev_1}}
  $\quad \tnext \in \{ \bot, \pea{\pip{\psi}}{\varphi} \}$
  \\[1em]
\end{center}
\begin{flushleft}
  Actions for~(\trei{}): \ \\
  \begin{tabular}[c]{lr}
    \begin{minipage}[c]{0.28\linewidth}
      \begin{eqnarray*}
        \tnext_1 & := & \tfean(\varphi)
        \\ \\
        \tbdia_1 & := & \tbl\big( \varphi, \tbdia \big)
      \end{eqnarray*}
    \end{minipage}
    \begin{minipage}[c]{0.72\linewidth}
      \begin{eqnarray*}
        \tuev(\chi_1, \chi_2) & := &
        \left\{
          \begin{array}{ll}
            \tuev_1(\varphi, \chi_2) & \text{ if } \chi_1 = \pea{\pip{\psi}}{\varphi}\\
            \tuev_1(\chi_1, \chi_2) & \text{ if } \chi_1 \in \Gamma\\
            \bot & \text{ otherwise}
          \end{array}
        \right .
      \end{eqnarray*}
    \end{minipage}
  \end{tabular}
\end{flushleft}

These rules just capture the transformations in Table~\ref{tab_alphabeta}
except for the histories.
Their choice of principal formula is free if~$\tnext = \bot$,
but is restricted to the formula in~$\tnext$ otherwise.
If the decomposition~$\chi$ of the principal \fea{}-formula is a \fean{}-formula,
we put~$\tnext_1$ of the child to be~$\chi$ to enforce that~$\chi$ is the principal formula of the child.
The actions for~$\tuev$ ensure
that~$\tuev(\chi_1, \chi_2)$, where~$\chi_1$ is the principal \fea{}-formula,
inherits its value from the corresponding \fea{}-formulae in the child:
\eg{} $\tuev(\pea{\psp{\alpha}{\beta}}{\varphi}, \chi_2) = \tuev_1(\pea{\alpha}{\pea{\beta}{\varphi}}, \chi_2)$
reverse-tracks the decomposition of~$\pea{\psp{\alpha}{\beta}}{\varphi}$ into~$\pea{\alpha}{\pea{\beta}{\varphi}}$.
Also, $\tuev(\chi_1, \chi_2)$ is only defined if~$\chi_1$ is in the parent.

\paragraph{Universal Branching ($\tbeta$) Rules.}

\begin{center} 
  \ruleone{(\tror{})}
  {\tnode{\por{\varphi_1}{\varphi_2} , \; \Gamma}{\tnext}{\tmrk, \tuev}}
  {\tnode{\varphi_1 , \; \Gamma}{\tnext}{\tmrk_1, \tuev_1}
    \mid
    \tnode{\varphi_2 , \; \Gamma}{\tnext}{\tmrk_2, \tuev_2}}
  $\quad \tnext = \bot$
  \\[1em]
  \ruleone{(\trai{})}
  {\tnode{\paa{\pip{\psi}}{\varphi} , \; \Gamma}{\tnext}{\tmrk, \tuev}}
  {\tnode{\pneg{\psi} , \; \Gamma}{\tnext}{\tmrk_1, \tuev_1}
    \mid
    \tnode{\varphi , \; \Gamma}{\tnext}{\tmrk_2, \tuev_2}}
  $\quad \tnext = \bot$
  \\[1em]
\end{center}
Action for~(\tror{}) and~(\trai{}) for~$i = 1, 2$:
$\tuev_i'(\chi_1, \chi_2) := 
\left\{
  \begin{array}{ll}
    \tuev_i(\chi_1, \chi_2) & \text{ if } \chi_1 \in \Gamma\\
    \bot & \text{ otherwise}
  \end{array}
\right .$

\begin{center}
  \ruleone{(\treu{})}
  {\tnode{\pea{\pup{\alpha_1}{\alpha_2}}{\varphi} , \; \Gamma}{\tnext, \tbdia}{\tmrk, \tuev}}
  {\begin{array}{l}
      \tnode{\pea{\alpha_1}{\varphi} , \; \Gamma}{\tnext_1, \tbdia_1}{\tmrk_1, \tuev_1} \; \mid 
      \tnode{\pea{\alpha_2}{\varphi} , \; \Gamma}{\tnext_2, \tbdia_2}{\tmrk_2, \tuev_2}
    \end{array}}
\end{center}
Side-condition for~(\treu{}): $\tnext \in \{ \bot, \pea{\pup{\alpha_1}{\alpha_2}}{\varphi} \}$
\begin{flushleft}
  Action for~(\treu{}) for $i = 1,2$:
  \begin{tabular}[c]{lr}
    \begin{minipage}[c]{0.26\linewidth}
      \begin{eqnarray*}
        \tnext_i & := & \tfean\big( \pea{\alpha_i}{\varphi} \big)
        \\ \\
        \tbdia_i & := & \tbl\big( \pea{\alpha_i}{\varphi}, \tbdia \big)
      \end{eqnarray*}
    \end{minipage}
    &
    \begin{minipage}[c]{0.74\linewidth}
      \begin{eqnarray*}
        \tuev_i'(\chi_1, \chi_2) & := &
        \left\{
          \begin{array}{ll}
            \tuev_i(\pea{\alpha_i}{\varphi}, \chi_2) & \text{ if } \chi_1 = \pea{\pup{\alpha_1}{\alpha_2}}{\varphi}\\
            \tuev_i(\chi_1, \chi_2) & \text{ if } \chi_1 \in \Gamma\\
            \bot & \text{ otherwise}
          \end{array}
        \right . \\
      \end{eqnarray*}
    \end{minipage}
  \end{tabular}
\end{flushleft}

\begin{center}
  \ruleone{(\trero{})}
  {\tnode{\pea{\prp{\alpha}}{\varphi} , \; \Gamma}{\tnext, \tbdia}{\tmrk, \tuev}}
  {\begin{array}{l}
      \tnode{\varphi , \; \Gamma}{\tnext_1, \tbdia_1}{\tmrk_1, \tuev_1} \; \mid 
      \tnode{\pea{\alpha}{\pea{\prp{\alpha}}{\varphi}} , \; \Gamma}%
      {\tnext_2, \tbdia_2}{\tmrk_2, \tuev_2}
  \end{array}}
\end{center}
Side-condition for~(\trero{}):
$\tnext \in \{ \bot, \pea{\prp{\alpha}}{\varphi} \} \xand \pea{\prp{\alpha}}{\varphi} \notin \tbdia$
\begin{flushleft}
  Action for~(\trero{}):
  \begin{tabular}[c]{lr}
    \begin{minipage}[c]{0.28\linewidth}
      \begin{eqnarray*}
        \tnext_1 & := & \tfean(\varphi)
        \\ \\
        \tbdia_1 & := & \tbl\big( \varphi, \{ \pea{\prp{\alpha}}{\varphi} \} \cup \tbdia \big)
      \end{eqnarray*}
    \end{minipage}
    &
    \begin{minipage}[c]{0.72\linewidth}
      \begin{eqnarray*}
        \tuev_1'(\chi_1, \chi_2) & := &
        \left\{
          \begin{array}{ll}
            \bot & \text{ if } \chi_1 = \chi_2 = \pea{\prp{\alpha}}{\varphi}\\
            \tuev_1(\varphi, \chi_2) & \text{ if } \chi_1 = \pea{\prp{\alpha}}{\varphi} \not= \chi_2\\
            \tuev_1(\chi_1, \chi_2) & \text{ if } \chi_1 \in \Gamma\\
            \bot & \text{ otherwise}
          \end{array}
        \right . \\
      \end{eqnarray*}
    \end{minipage}
  \end{tabular}
  \begin{tabular}[c]{lr}
    \begin{minipage}[c]{0.26\linewidth}
      \begin{eqnarray*}
        \tnext_2 & := & \tfean\big( \pea{\alpha}{\pea{\prp{\alpha}}{\varphi}} \big)
        \\[1em]
        \\
        \tbdia_2 & := & \tbl\big( \pea{\alpha}{\pea{\prp{\alpha}}{\varphi}}, \{ \pea{\prp{\alpha}}{\varphi} \} \cup \tbdia \big)
      \end{eqnarray*}
    \end{minipage}
    &
    \begin{minipage}[c]{0.74\linewidth}
      \begin{eqnarray*}
        \tuev_2'(\chi_1, \chi_2) & := &
        \left\{
          \begin{array}{ll}
            \tuev_2(\pea{\alpha}{\pea{\prp{\alpha}}{\varphi}}, \chi_2) & \text{ if } \chi_1 = \pea{\prp{\alpha}}{\varphi}\\
            \tuev_2(\chi_1, \chi_2) & \text{ if } \chi_1 \in \Gamma\\
            \bot & \text{ otherwise}
          \end{array}
        \right .
      \end{eqnarray*}
    \end{minipage}
  \end{tabular}
\end{flushleft}

The \trero{}-rule captures the fix-point nature of the \fear{}-formulae
according to Prop.~\ref{prop_axioms}
as long as the principal formula is not blocked via~$\tbdia$.
The choice of the principal formulae in the first child is either free
if~$\varphi$ is not a \fean{}-formula
or is~$\varphi$ if~$\varphi$ is a \fean{}-formula.
In the latter case we also block the regeneration of~$\pea{\prp{\alpha}}{\varphi}$
and thus avoid an ``at a world'' cycle by putting~$\pea{\prp{\alpha}}{\varphi}$ into~$\tbdia_1$.
The right child is treated similarly
but uses~$\pea{\alpha}{\pea{\prp{\alpha}}{\varphi}}$ instead of~$\varphi$.
\begin{flushleft}
  Actions for all $\tbeta$-rules: 
  \begin{eqnarray*}
    \tmrk & := &
    \left\{
      \begin{array}{ll}
        \ttrue & \text{ if } \tmrk_1 = \ttrue \xand \tmrk_2 = \ttrue\\
        \tfalse & \text{ if } \tmrk_1 = \tfalse \xor \tmrk_2 = \tfalse\\
        \tloop & \text{ otherwise }
      \end{array}
    \right .
    \\
    \tmin(f, g)(\chi_1, \chi_2) & := &
    \left\{
      \begin{array}{l}
        \bot \qquad \text{ if } f(\chi_1, \chi_2) = \bot \text{ or } g(\chi_1, \chi_2) = \bot\\
        \min(f(\chi_1, \chi_2), g(\chi_1, \chi_2)) \qquad \text{ otherwise}
      \end{array}
    \right .
    \\
    \tuev & := &
    \left\{
      \begin{array}{ll}
        \tuev_{\bot} & \text{ if } \tmrk \not= \tfalse\\
        \tuev_1' & \text{ if } \tmrk_1 = \tfalse \not= \tmrk_2\\
        \tuev_2' & \text{ if } \tmrk_1 \not= \tfalse = \tmrk_2\\
        \tmin(\tuev_1', \tuev_2') & \text{ if } \tmrk_1 = \tfalse = \tmrk_2
      \end{array}
    \right .
  \end{eqnarray*}
\end{flushleft}

The intuitions are:
\begin{description}
\item[\rm $\tuev_i'$:] the definitions of~$\tuev_i'$ ensure
  that the pairs~$(\chi_1, \chi_2)$, where~$\chi_1$ is the principal \fea{}-formula,
  get the values from their corresponding \fea{}-formulae in the children.
  In the \trero{}-rule, a special case sets
  the value of~$\tuev_1'(\chi_1, \chi_2)$ to~$\bot$
  if~$\chi_1$ and~$\chi_2$ are equal
  to the principal formula~$\pea{\prp{\alpha}}{\varphi}$ of this rule
  since the eventuality~$\pea{\prp{\alpha}}{\varphi}$
  is no longer unfulfilled as the left child fulfils it.
  Note that~$\tuev'(\chi_1, \chi_2)$ is only defined
  if~$\chi_1$ is in the parent.
\item[\rm $\tmin$:] the definition of~$\tmin$ ensures
  that we take the minimum of~$f(\chi_1, \chi_2)$ and~$g(\chi_1, \chi_2)$
  only when both functions are defined for~$(\chi_1, \chi_2)$.
\item[\rm $\tuev$:] if~$\tmrk \neq \tfalse$,
  the~$\tuev$ is irrelevant,
  so we arbitrarily set it as undefined.
  If only one child has~$\tmrk = \tfalse$, we take its~$\tuev'$.
  If both children have~$\tmrk = \tfalse$,
  we take the minimum value of entries
  that are defined in~$\tuev_1'$ and~$\tuev_2'$.
\end{description}

All previous rules modify existing $\tuev$-entries, but never
create new ones.
The next rule is the only rule that creates $\tuev$-entries (by identifying loops).

\paragraph{Existential Branching Rule.}

\begin{center}
  \ruleone{(\trea{})}
  {\begin{array}{l}
      \pea{a_1}{\varphi_1} , \dotsc, \pea{a_n}{\varphi_n} , \;
      \pea{a_{n+1}}{\varphi_{n+1}} , \dotsc, \pea{a_{n+m}}{\varphi_{n+m}} , \;
      \paa{-}{\Delta} , \; \Gamma\\
      :: \thc, \tnext, \tbdia, \tbbox :: \tmrk, \tuev
    \end{array}}
  {\begin{array}{l}
      \varphi_1 , \; \Delta_1 :: \thc_1, \tnext_1, \tbdia_1, \tbbox_1 \\
      :: \tmrk_1, \tuev_1
    \end{array}
    \mid \dotsm \mid
    \begin{array}{l}
      \varphi_n , \; \Delta_n :: \thc_n, \tnext_n, \tbdia_n, \tbbox_n \\
      :: \tmrk_n, \tuev_n
    \end{array}}
\end{center}
\noindent where:
\renewcommand{\theenumi}{(\arabic{enumi})}
\renewcommand{\labelenumi}{\theenumi}
\begin{enumerate}
\item\label{enum_one} $n + m \geq 0$
\item\label{enum_two} $\Gamma \subseteq \big( \proptn \cup \{ \pnot{q} \mid q \in \proptn \} \big)$
\item\label{enum_thr} $\paa{-}{\Delta} \subseteq \big\{ \paa{a}{\psi} \mid a \in \act \xand \psi \in \fml \big\}$
\item\label{enum_fou} $\Delta_i := \{ \psi \mid \paa{a_i}{\psi} \in \paa{-}{\Delta} \}$ for $i = 1, \dotsc, n$
\item\label{enum_fiv} $\forall p \in \proptn.\: \{p, \pnot{p}\} \not\subseteq \Gamma$
\item\label{enum_six}
  $\forall i \in \{1, \dotsc, n\}.\: \forall j \in \{1, \dotsc, \llen(\thc) \}.\: \big( \varphi_i, \{ \varphi_i \} \cup \Delta_i \big) \neq \thc[j]$
\item\label{enum_sev}
  $\forall k \in \{n+1, \dotsc, n+m\}.\: \exists j \in \{1, \dotsc, \llen(\thc) \}.\: \big( \varphi_k, \{ \varphi_k \} \cup \Delta_k \big) = \thc[j]$
\end{enumerate}
Actions for~(\trea):
\begin{array}[t]{rl}
  \text{for } i = 1, \dotsc, n:\;
  &
  \thc_i := \thc \;@\; \big[ \big( \varphi_i, \{ \varphi_i \} \cup \Delta_i \big) \big],\\
  &
  \tnext_i := \tfean(\varphi_i), \quad
  \tbdia_i := \emptyset, \quad
  \tbbox_i := \emptyset
\end{array}
\begin{eqnarray*}
  \tmrk & := &
  \left\{
    \begin{array}{ll}
      \ttrue & \text{ if}
      \begin{array}[t]{l}
        \exists i \in \{1, \dotsc , n\}.\: \tmrk_i \not= \tfalse \xor\\
        \big( \exists \psi \in \fmler.\: \varphi_i \in \ppre{\psi} \xand\\
        \bot \neq \tuev_i(\varphi_i, \psi) > \llen(\thc) \big)
      \end{array}\\
      \tfalse & \text{ otherwise }
    \end{array}
  \right .
  \\[1em]
  \tuev_k(\cdot, \cdot) & := & j \in \{1, \dotsc, \llen(\thc)\}
  \text{ such that } \big( \varphi_k, \{ \varphi_k \} \cup \Delta_k \big) = \thc[j] \\
  & & \text{for } k = n+1, \dotsc ,n+m
  \\[1em]
  \tuev(\chi_1, \chi_2) & := &
  \left\{
    \begin{array}{ll}
      \tuev_i(\varphi_i, \chi_2) & \text{ if}
      \begin{array}[t]{l}
        \tmrk = \tfalse \xand \chi_2 \in \fmler \xand \chi_1 \in \ppre{\chi_2}\\
        \xand \chi_1 = \pea{a_i}{\varphi_i} \text{ for an } i \in \{ 1, \dotsc , n+m \}\\
      \end{array}\\
      \bot & \text{ otherwise }
    \end{array}
  \right .
\end{eqnarray*}

Some intuitions are in order:
\begin{description}
\item[\rm \ref{enum_one}]
  If~$n = 0$, the application of the rule generates no new nodes
  and~$\tmrk$ vacuously evaluates to~$\tfalse$.
  If~$m = n = 0$, we additionally have~$\tuev := \tuev_{\bot}$.
\item[\rm \ref{enum_two}] The set~$\Gamma$ contains only propositional atoms or their negations.
\item[\rm \ref{enum_thr}] The set~$\paa{-}{\Delta}$ contains only formulae of the type~$\paa{a}\varphi$.
  Thus~\ref{enum_two} and~\ref{enum_thr} imply
  that the \trea{}-rule is applicable only
  if the node contains no $\talpha$- or $\tbeta$-formulae.
\item[\rm \ref{enum_fou}] The set~$\Delta_i$ contains all formulae
  that must belong to the $i^{\rm th}$~child, which fulfils~$\pea{a_i}{\varphi_i}$,
  so that we can build a Hintikka structure later on.
\item[\rm \ref{enum_fiv}] The node must not contain a contradiction.
\item[\rm \ref{enum_six}] If~$n > 0$,
  then each~$\pea{a_i}{\varphi_i}$ for~$1 \leq i \leq n$ is not ``blocked'' by an ancestor
  and has a child containing the formula set~$\varphi_i \cup \Delta_i$
  thereby generating the required successor for~$\pea{a_i}{\varphi_i}$.
  Note that~$\llen(\thc)$ denotes the length of~$\thc$.
\item[\rm \ref{enum_sev}] If~$m > 0$,
  then each~$\pea{a_k}{\varphi_k}$ for~$n+1 \leq k \leq n+m$ is ``blocked''
  from creating its required child~$\{ \varphi_k \} \cup \Delta_k$
  because some ancestor does the job.
  This ancestor must not only consist of the formulae~$\{ \varphi_k \} \cup \Delta_k$
  but it must also have been created to fulfil~$\pea{a}{\varphi_k}$ for some~$a \in \act$.
  Note that the values~$a_k$ and~$a$ are ignored
  when looking for loops
  since we are interested only in the contents of the required child.
\item[\rm $\thc_i$:] is the~$\thc$ of the parent
  extended with an extra entry to record the ``history''
  of worlds created on the path from the root down to the $i^{\rm th}$~child
  using ``@'' as list concatenation.
  Note that we store a \emph{pair}~$(\varphi_k, \varphi_k \cup
  \Delta_k)$, not just~$\varphi_k \cup \Delta_k$.
  That is, we remember that the node~$\varphi_k \cup \Delta_k$ was created
  to fulfil~$\pea{a}{\varphi_k}$ for some~$a \in \act$.
\item[\rm $\tmrk$:] the parent is unsatisfiable
  if some child has~$\tmrk \neq \tfalse$.
  But it is also unsatisfiable if some child, say the~$i^{\rm th}$,
  and some eventuality~$\pea{\prp{\alpha}}{\chi}$ in it ``loops lower''
  because~$\varphi_i \in \ppre{\pea{\prp{\alpha}}{\chi}}$
  and~$\tuev_i(\varphi_i, \pea{\prp{\alpha}}{\chi})$ is defined and greater than the length of the current~$\thc$.
  Intuitively, the latter tells us
  that the eventuality~$\pea{\prp{\alpha}}{\chi}$
  occurs in the sub-tableau rooted at the parent
  but cannot be fulfilled.
\item[\rm $\tuev_k$:] for~$n+1 \leq k \leq n+m$, the $k^{\rm th}$~child is blocked
  by a higher (proxy) child.
  For every such~$k$
  we set~$\tuev_k$ to be the \emph{constant} function
  which maps every formula-pair to the \emph{level}~$j$ of its proxy child.
  This is just a temporary function
  used to define~$\tuev$ as explained next.
  The blocking child itself must have been created
  to fulfil a \fea{}-formula~$\pea{a'}{\varphi_k}$,
  as indicated by the first component of~$\thc[j]$.
\item[\rm $\tuev(\chi_1, \chi_2)$:]
  If~$\tmrk = \ttrue$ then~$\tuev$ is undefined everywhere.
  Else,
  for each~$\chi_1 = \pea{a_i}{\varphi_i}$ with~$i \in \{ 1, \dotsc , n+m \}$,
  and each~$\chi_2$ with~$\pea{a_i}{\varphi_i} \in \ppre{\chi_2}$,
  we take~$\tuev(\pea{a_i}{\varphi_i}, \chi_2)$
  from the formulae-pair~$(\varphi_i, \chi_2)$ of the corresponding (real) child
  if~$\pea{a_i}{\varphi_i}$ is ``unblocked'',
  or set it to the level of the proxy child higher in the branch
  if it is ``blocked''.
  For all other formulae-pairs, $\tuev$ is undefined.
  The intuition is that a defined~$\tuev(\chi_1, \chi_2)$ flags
  a ``loop'' which starts at the parent
  and eventually ``loops'' up to some blocking proxy.
  The value of~$\tuev(\chi_1, \chi_2)$ tells us the level of the proxy
  because we cannot classify this ``loop'' as ``good'' or ``bad''
  until we backtrack to that level.
  The~$\tuev$ of each~$\pea{a_i}{\varphi_i}$ is taken from \emph{the} child
  created specifically to contain~$\varphi_i$,
  a fact which is vital in the proofs.
\item[\rm $\tbdia_i, \tbbox_i, \tnext_i$:]
  each child has no blocked diamond- or box-formulae,
  and its principal formula is determined by the form of $\varphi_i$.
\end{description}

The \trea{}- and \trid{}-rules are mutually exclusive via their
side-conditions.  Our rules are designed so that at least one rule is
applicable to any node. As shown in the next section, we need to build
only one fully expanded tableau, hence if multiple rules are
applicable to a node, the choice of rule is immaterial.
Of course, in our implementation, we give priority to the \trid{}-rule
since it may close a branch sooner. Other heuristics, like preferring
linear rules over branching rules, are also useful.

\subsection{Termination, Soundness, and Completeness}

\begin{definition}
  Let~$x = \tnode{\Gamma}{\thc, \tnext, \tbdia, \tbbox}{\tmrk, \tuev}$ be a tableau node,
  $\varphi$ a formula, and~$\Delta$ a set of formulae.
  We write~$\varphi \in x$ $[\Delta \subseteq x]$ to mean~$\varphi \in \Gamma$ $[\Delta \subseteq \Gamma]$.
  The parts of~$x$
  are written as~$\thc_x$, $\tnext_x$, $\tbdia_x$, $\tbbox_x$, $\tmrk_x$, and~$\tuev_x$.
  Node~$x$ is \emph{closed} iff~$\tmrk_x = \ttrue$,
  \emph{open} iff~$\tmrk_x = \tfalse$, and \emph{barred} iff~$\tmrk_x = \tloop$.
\end{definition}

\begin{definition}
  Let~$x$ be a \trea{}-node in a tableau~$T$
  (i.e.\ a \trea{}-rule was applied to~$x$).
  Then~$x$ is also called a \emph{state} and
  the children of~$x$ are called \emph{core-nodes}.
  Using the notation of the \trea{}-rule,
  a formula~$\pea{a_i}{\varphi_i} \in x$ is \emph{blocked} iff~$n+1 \leq i \leq n+m$.
  For every not blocked~$\pea{a_i}{\varphi_i} \in x$,
  the \emph{successor} of~$\pea{a_i}{\varphi_i}$ is the $i^{\mathrm{th}}$~child of the \trea{}-rule.
  For every blocked~$\pea{a_i}{\varphi_i} \in x$
  there exists a unique core-node~$y$ on the path from the root of~$T$ to~$x$
  such that~$\{ \varphi_i \} \cup \Delta_i$ is the set of formulae of~$y$,
  and~$y$ is the successor of a formula~$\pea{a'}{\varphi_i}$
  in the parent of~$y$.
  We call~$y$ the \emph{virtual successor} of~$\pea{a_i}{\varphi_i}$,
  and also call the formula~$\varphi_i$ in the (possibly virtual) successor of~$\pea{a_i}{\varphi_i}$
  a \emph{core-formula}.
\end{definition}
A state is another term for a \trea{}-node
but a core-node can be any type of node
(even a state).
A state arises from a core-node
by $\talpha$- and $\tbeta$-rules.
Note that the core-formula in a core-node~$y$ is well-defined and unique:
if~$x_1$ and~$x_2$ are states
and~$y$ is the (possibly virtual) successor of~$\pea{a_1}{\varphi_1} \in x_1$ and~$\pea{a_2}{\varphi_2} \in x_2$,
then~$\varphi_1 = \varphi_2$.

Let~$\phi$ be a formula in negation normal form,
and~$T$ an expanded tableau with root~$r = \tnode{\{\phi\}}{[], \bot, \emptyset, \emptyset}{\tmrk, \tuev}$
with~$\tmrk$ and~$\tuev$ determined by $r$'s children.
\begin{theorem}
  \label{thm_term}
  $T$ is a finite tree.
\end{theorem}
\begin{theorem}
  \label{theo_correctness}
  If the root~$r \in T$ is open, there is a Hintikka structure for~$\phi$.
\end{theorem}
\begin{theorem}
  \label{theo_completeness}
  If the root~$r \in T$ is not open then~$\phi$ is not satisfiable.
\end{theorem}
\begin{theorem}
  \label{theo_complexity}
  If~$|\phi| = n$, our procedure has worst-case time complexity in~$O(2^{2^{n}})$.
\end{theorem}
The length of a branch in a tableau is bounded, essentially
by the number of core-nodes on that branch.
The number of core-nodes itself is bounded, essentially
by the cardinality of the power set of the set~$\pcl{\phi}$ of all formula
that can appear in the tableau.
The size of~$\pcl{\phi}$ is polynomial in~$n$,
hence the length of a branch is in~$O(2^n)$.
Thus the overall (worst case) number of nodes in a tableau is in~$O(2^{2^n})$.

\subsection{Fully Worked Examples}

The first simple example illustrates
how the procedure avoids infinite loops due to ``at a world'' cycles
by blocking $\pea{\prp{\alpha}}{\varphi}$- and $\paa{\prp{\alpha}}{\varphi}$-formulae from regenerating.
The formula~$\extst$ is obviously not satisfiable.
Hence, any expanded tableau with root~$\extst$ should not be open.
Figure~\ref{fig_ex2} shows such a tableau
where each node is classified as a $\rho$-node
if rule~$\rho$ is applied to that node in the tableau.

The initial formula~$\extst$ in node~(1) 
is decomposed into a $\tbeta_1$-child~$\pand{p}{\pnot{p}}$
and a $\tbeta_2$-child~$\pea{\pip{q}}{\extst}$
according to the \trero{}-rule.
The formula~$\pand{p}{\pnot{p}}$ in node~(2) is then decomposed
according to the \trand{}-rule and node~(3) is marked as closed
because it contains a contradiction.
Node~(2) inherits the status from node~(3) unchanged 
according to the $\talpha$-rules
and, thus, is closed too.

Because the $\tbeta_2$-formula~$\pea{\pip{q}}{\extst}$ is a \fean{}-formula,
the \trero{}-rule puts this formula into its~$\tnext_2$,
the~$\tnext$-value of node~(4),
and thus forces node~(4) to have~$\pea{\pip{q}}{\extst}$ as its principal formula.
For the same reason, 
the \trero{}-rule puts its own principal formula~$\extst$
into its~$\tbdia_2$, the $\tbdia$-value of node~(4).
Hence node~(4) decomposes~$\pea{\pip{q}}{\extst}$ according to the \trei{}-rule.
Again, the resulting node~(5) is forced to have~$\extst$ as its principal formula
via its $\tnext$-value, and gets its~$\tbdia$-value unchanged from node~(4).

Node~(5) has the same principal formula as node~(1),
so applying the \trero{}-rule to node~(5) would cause
the procedure to enter an ``at a world'' (infinite) cycle.
Because the history~$\tbdia$ of node~(5) contains~$\extst$,
the \trero{}-rule is blocked on node~(5),
but the \trert{}-rule is not.
Hence the branch is terminated and the status of node~(5) is set to $\tloop$
(thereby avoiding the ``at a world'' cycle).

Node~(4) inherits the status from node~(5) unchanged
and node~(1) is marked $\tloop$ also
according to the definition of~$\tmrk$ in the $\tbeta$-rules.
Therefore the tableau is not open.
Note that the variable~$\tuev$ does not play a role in this example
as it is irrelevant for nodes that are closed or barred.
\begin{figure}
  \begin{center}
    \begin{math}
      \xymatrix{
        *+[F]{
          \begin{tabular}{c}
            (2) \hfill \trand{}-node\\
            $\pand{p}{\pnot{p}}$\\ 
            $:: [], \bot, \emptyset, \emptyset :: \ttrue, \tuev_{\bot}$
          \end{tabular}
        }
        \ar[d]_-{\talpha}
        &
        *+[F=]{
          \begin{tabular}{c}
            (1) \hfill \trero{}-node\\
            $\extst$\\
            $:: [], \bot, \emptyset, \emptyset :: \tloop, \tuev_{\bot}$
          \end{tabular}
        }
        \ar[l]^-{\tbeta_1}
        \ar[d]_-{\tbeta_2}
        \\
        *+[F]{
          \begin{tabular}{c}
            (3) \hfill \trid{}-node\\
            $p \ , \ \pnot{p}$\\
            $:: [], \bot, \emptyset, \emptyset :: \ttrue, \tuev_{\bot}$
          \end{tabular}
        }
        &
        *+[F]{
          \begin{tabular}{c}
            (4) \hfill \trei{}-node\\
            $\pea{\pip{q}}{\extst}$\\
            $\begin{array}{l}
              :: [], \pea{\pip{q}}{\extst}, \{ \extst \}, \emptyset\\
              :: \tloop, \tuev_{\bot}
            \end{array}$
          \end{tabular}
        }
        \ar[d]_-{\talpha}
        \\
        &
        *+[F]{
          \begin{tabular}{c}
            (5) \hfill \trert{}-node\\
            $q \ , \ \extst$\\
            $\begin{array}{l}
            :: [], \extst, \{ \extst \}, \emptyset\\
            :: \tloop, \tuev_{\bot}
            \end{array}$
          \end{tabular}
        }
      }
    \end{math}
  \end{center}
  \caption[]{A first example: a closed tableau for~$\extst$}
  \label{fig_ex2}
\end{figure}

The second example demonstrates the role of~$\tuev$.
The formula~$\pimp{\exasp}{\paa{\prp{(\psp{a}{a})}}{p}}$ is valid.
Hence, its negation~$\phi := \pand{\exasp}{\exaasp}$,
which is already in negation normal form, is unsatisfiable
and the root of any expanded tableau for~$\phi$ should not be open.
Figure~\ref{fig_ex1} shows such a tableau.
The unlabelled edges in Fig.~\ref{fig_ex1} link states to core-nodes.
We omit the histories~$\tbdia$ and~$\tbbox$ 
as they do not play an important role in this example.
Each partial function~$UEV_i$ maps the formula-pair~$(\psi_i, \chi_i)$
in Table~\ref{tab_uev} to~1
and is undefined otherwise as explained below.
The histories are~$HCR_1 := [(\varphi_1, \Delta_1)]$
where $\varphi_1 := \pea{a}{\exaasp}$ and $\Delta_1 := \{ \exasp, \pea{a}{\exaasp} \}$
and~$HCR_2 := HCR_1 @ [(\varphi_2, \Delta_2)]$ where $\varphi_2 := \exaasp$ and $\Delta_2 := \{ \exasp, \exaasp \}$.
\begin{table}[t]
  \caption{Definitions for the example in Fig.~\ref{fig_ex1}}
  \label{tab_uev}
  \begin{center}
    \begin{tabular}{|c|c|c|c|c|}
      \hline
      $UEV_i$ & $i = 1$ & $i = 2$ & $i = 3$ & $i = 4$ 
      \\ \hline
      $\psi_i$ 
      & $\pea{a}{\pea{a}{\exaasp}}$ 
      & $\pea{\psp{a}{a}}{\exaasp}$ 
      & $\exaasp$ 
      & $\pea{a}{\exaasp}$ 
      \\ \hline
      $\chi_i$
      & $\exaasp$ & $\exaasp$ & $\exaasp$ & $\exaasp$
      \\ \hline
    \end{tabular}
  \end{center}
\end{table}

The dotted frame at~(7a) indicates
that its child, an \trid{}-node,
is not shown due to space restrictions.
Thus the marking of the nodes~(3a) and~(7a) in Fig.~\ref{fig_ex1} with~$\ttrue$ is straightforward.
The leaf~(9) is a \trea{}-node,
but it is ``blocked'' from creating its successor containing~$\Delta := \{ \exasp, \pea{a}{\exaasp} \}$
because there is a~$j \in \Nat$
such that $\thc_9[j] = HCR_2[j] = (\pea{a}{\exaasp}, \Delta)$: namely~$j=1$.
Thus the \trea{}-rule computes $UEV_1(\pea{a}{\varphi_1}, \exaasp) = 1$ as stated above
and also puts $\tmrk_9 := \tfalse$.
As node~(7a) is closed,
nodes~(8), (7b), (7), (6), and~(5)
inherit their functions~$UEV_i$ from their open children
via the corresponding $\talpha$- and $\tbeta$-rules.

The crux of our method occurs at node~(4),
a \trea{}-node with $\thc_4 = []$ and hence~$\llen(\thc_4) = 0$.
The \trea{}-rule thus finds a child node~(5)
and a pair of formulae~$(\psi, \chi) := (\pea{a}{\exaasp}, \exaasp)$
where~$\psi$ is a core-formula, $\psi \in \ppre{\chi}$,
and $1 = UEV_4(\psi, \chi) = \tuev_5(\psi, \chi) > \llen(\thc_4) = 0$.
Thus node~(4) ``sees'' a child~(5) that ``loops lower'',
meaning that node~(5) is the root of an ``isolated'' subtree
which fails to fulfil its eventuality~$\exaasp$.
The \trea{}-rule marks~(4) as closed via~$\tmrk_4 = \ttrue$.
The propagation of $\ttrue$ to the root is simple.

What if the omitted child of~(7a), and hence~(7a) itself, had been open?
Then~$UEV_3$ in~(7) would be undefined everywhere via the
\trero{}-rule, regardless of~$\tuev_{7\mathrm{b}}$.
Thus~$\exaasp$ in~(7) would be fulfilled via the $\tbeta_1$-child~(7a).
Hence~$UEV_4$ would be undefined everywhere,
and node~(4) would not be closed.
\begin{figure}
  \begin{center}
    \begin{math}
      \xymatrix{
        *+[F=]{
          \begin{tabular}{c}
            (1) \hfill \trand{}-node\\
            $\pand{\exasp}{\exaasp}$\\
            $:: [], \bot :: \ttrue, \tuev_{\bot}$
          \end{tabular}
        }
        \ar[r]^-{\talpha}
        &
        *+[F]{
          \begin{tabular}{c}
            (2) \hfill \trar{}-node\\
            $\exasp \ , \ \exaasp$\\
            $:: [], \bot :: \ttrue, \tuev_{\bot}$
          \end{tabular}
        }
        \ar[d]_-{\talpha}
        \\
        *+[F]{
          \begin{tabular}{c}
            (3a) \hfill \trid{}-node\\
            $p \ , \ \paa{a}{\exasp} \ , \ \pnot{p}$\\
            $:: [], \bot :: \ttrue, \tuev_{\bot}$
          \end{tabular}
        }
        &
        *+[F]{
          \begin{tabular}{c}
            (3) \hfill \trero{}-node\\
            $p \ , \ \paa{a}{\exasp} \ , \ \exaasp$\\
            $:: [], \bot :: \ttrue, \tuev_{\bot}$
          \end{tabular}
        }
        \ar[l]_-{\tbeta_1}
        \ar[d]_-{\tbeta_2}
        \\
        *+[F]{
          \begin{tabular}{c}
            (4) \hfill \trea{}-node\\
            $p \ , \ \paa{a}{\exasp} \ , \ \pea{a}{\pea{a}{\exaasp}}$\\
            $:: [], \bot :: \ttrue, \tuev_{\bot}$
          \end{tabular}
        }
        \ar[d]
        &
        *+[F]{
          \begin{tabular}{c}
            (3b) \hfill \tres{}-node\\
            $p \ , \ \paa{a}{\exasp} \ , \ \pea{\psp{a}{a}}{\exaasp}$\\
            $:: [], \pea{\psp{a}{a}}{\exaasp} :: \ttrue, \tuev_{\bot}$
          \end{tabular}
        }
        \ar[l]_-{\talpha}
        \\
        *+[F]{
          \begin{tabular}{c}
            (5) \hfill \trar{}-node\\
            $\exasp \ , \ \pea{a}{\exaasp}$\\
            $:: HCR_1, \bot :: \tfalse, UEV_4$
          \end{tabular}
        }
        \ar[r]^-{\talpha}
        &
        *+[F]{
          \begin{tabular}{c}
            (6) \hfill \trea{}-node\\
            $p \ , \ \paa{a}{\exasp} \ , \ \pea{a}{\exaasp}$\\
            $:: HCR_1, \bot :: \tfalse, UEV_4$
          \end{tabular}
        }
        \ar[d]
        \\
        *+[F.]{
          \begin{tabular}{c}
            (7a) \hfill \trar{}-node\\
            $\exasp \ , \ \pnot{p}$\\
            $:: HCR_2, \bot :: \ttrue, \tuev_{\bot}$
          \end{tabular}
        }
        &
        *+[F]{
          \begin{tabular}{c}
            (7) \hfill \trero{}-node\\
            $\exasp \ , \ \exaasp$\\
            $:: HCR_2, \exaasp :: \tfalse, UEV_3$
          \end{tabular}
        }
        \ar[l]_-{\tbeta_1}
        \ar[d]_-{\tbeta_2}
        \\
        *+[F]{
          \begin{tabular}{c}
            (8) \hfill \trar{}-node\\
            $\exasp \ , \ \pea{a}{\pea{a}{\exaasp}}$\\
            $:: HCR_2, \bot :: \tfalse, UEV_1$
          \end{tabular}
        }
        \ar[d]_-{\talpha}
        &
        *+[F]{
          \begin{tabular}{c}
            (7b) \hfill \tres{}-node\\
            $\exasp \ , \ \pea{\psp{a}{a}}{\exaasp}$\\
            $:: HCR_2, \pea{\psp{a}{a}}{\exaasp} :: \tfalse, UEV_2$
          \end{tabular}
        }
        \ar[l]_-{\talpha}
        \\
        *+[F]{
          \begin{tabular}{c}
            (9) \hfill \trea{}-node\\
            $p \ , \ \paa{a}{\exasp} \ , \ \pea{a}{\pea{a}{\exaasp}}$\\
            $:: HCR_2, \bot :: \tfalse, UEV_1$
          \end{tabular}
        }
        \ar[r]
        &
        *+[F]{
          \begin{tabular}{c}
            blocked by node~(5)
          \end{tabular}
        }
      }
    \end{math}
  \end{center}
  \caption[]{A second example: a closed tableau for~$\pand{\exasp}{\exaasp}$}
  \label{fig_ex1}
\end{figure}

\section{Conclusion and Further Work}

We have given a sound, complete and terminating procedure
for checking \pdl{}-satisfiability.
Unfortunately, its worst-case time-complexity is in 2EXPTIME rather than in EXPTIME,
thus our procedure is sub-optimal.
We now outline some further practical and theoretical work
which may eliminate this disadvantage.

First, we believe that a small refinement of our histories will allow
our calculus to classify a loop as ``bad'' or ``good'' at the looping leaf,
as is done by Baader's procedure~\cite{baader-augmenting-transitive-closure},
but with no pre-computation of automata.
Thus it should be possible to extend DLP to handle our method.
Further experimental work is required to determine
if such an extension will remain practical.

Second, recent work has shown that global caching
can indeed deliver optimality of tableau procedures
soundly~\cite{gore-nguyen-exptime-alc}.
The histories used in our calculus make it harder to extend sound global caching to it
since nodes are now sensitive to their context in the tree under construction.
Further theoretical work is required
to extend sound global caching to handle such context sensitivity.


\newpage

\section*{Appendix: Termination, Soundness and Completeness}

\begin{definition}
  Let~$G = (W, R)$ be a directed graph
  (\eg{} a tableau
  where~$R$ is just the child-of relation between nodes).
  A \emph{path}~$\pi$ in~$G$
  is a finite or infinite sequence~$x_0, x_1, x_2, \dotsc$ of nodes in~$W$
  such that~$\prel{x_i}{R}{x_{i+1}}$ for all~$x_i$
  except the last node if~$\pi$ is finite.
\end{definition}

\paragraph{\bf Termination}

\noindent{}\textbf{Theorem~\ref{thm_term}\ }
\emph{$T$ is a finite tree.}
\begin{proof*}{Proof Sketch}
  It is obvious that~$T$ is a tree
  and that every node in~$T$ can contain only formulae
  from the negation normal form analogue~$\pcl{\phi}$
  of the Fisher-Ladner closure~\cite{fischer-ladner-dynamic}.
  The definition of~$\pcl{\phi}$ has been omitted to save space,
  but~$\pcl{\phi}$ is finite.
  Hence there are only a finite number of different sets
  that can be assigned to nodes,
  in particular core-nodes,
  and the number of pairs~$(\varphi, \Delta)$
  with~$\varphi \in \Delta \subseteq \pcl{\phi}$ is finite.
  As each core-node is assigned such a pair
  and the \trea{}-rule ensures
  core-nodes on a branch possess different pairs,
  the number of core-nodes on a branch is finite.

  It is not obvious
  that the number of nodes between consecutive core-nodes on a branch is finite
  since $\pea{\prp{\alpha}}{}$-
  and $\paa{\prp{\alpha}}{}$-formulae like~$\pea{\prp{\prp{a}}}{\varphi}$
  can ``regenerate'' on a branch without passing a core-node
  (\eg{}~$\pea{\prp{\prp{a}}}{\varphi} \pzz \pea{\prp{a}}\pea{\prp{\prp{a}}}{\varphi} \pzz
  \pea{\prp{\prp{a}}}{\varphi}$).
  However, it is relatively easy to see
  that formulae of the form~$\pea{\prp{\alpha}}{\varphi}$ or~$\paa{\prp{\alpha}}{\varphi}$
  are the only potential ``troublemakers'' between two states.
  For formulae of the form~$\paa{\prp{\alpha}}{\varphi}$
  regeneration between two core-nodes is clearly ruled out
  by the history~$\tbbox$ and the \trar{}-rule.
  For formulae of the form~$\pea{\prp{\alpha}}{\varphi}$,
  the job is done by the history~$\tbdia$ and the \trero{} and \trert{}-rules.
  In the latter case, it is crucial
  that the procedure chooses the decomposition of a principal \fean{}-formula
  as the principal formula of the child,
  provided that the decomposition is also a \fean{}-formula.

  As the number of nodes between two core-nodes is finite,
  and there are only finitely many core-nodes on any branch,
  all branches in~$T$ are finite.
  Every node has finite degree
  so K\"onig's lemma completes the proof.
\end{proof*}

\paragraph{\bf Soundness}

\noindent{}\textbf{Theorem~\ref{theo_correctness}\ }
\emph{If the root~$r \in T$ is open, there is a Hintikka structure for~$\phi$.}
\begin{proof}
  By construction, $T$ is a finite tree.
  Let~$\tm$ (``p'' for pruned)
  be the subgraph that consists of all nodes~$x$ having the following property:
  there is a path of open nodes from~$r$ to~$x$ inclusive.
  The edges of~$\tm$ are exactly the edges of~$T$ that connect two nodes in~$\tm$.
  Clearly, $\tm$ is also a finite tree with root~$r$.
  Intuitively, $\tm$ is the result of pruning all subtrees of~$T$ that have a closed or barred root.

  Next, we extend~$\tm$ to a finite cyclic tree~$\tl$ (``l'' for looping)
  by doing the following for every state~$x$:
  for every formula~$\pea{a}{\varphi} \in x$ having a \emph{virtual} successor~$y$,
  which must lie on the path from~$r$ to~$x$,
  we add the edge~$(x, y)$ to~$\tl$.
  Theses new edges are called \emph{backward edges}.
  Note that as \trid{}-nodes are closed by construction of~$T$,
  all leaves of~$\tm$ must be states where all \fea{}-formulae (if any) are blocked.
  Hence every formula~$\pea{a}{\varphi}$ of every leaf has a virtual successor.

  Finally, following Ben-Ari et al.~\cite{ben-ari-pnueli-manna-branching},
  the cyclic tree~$\tl$ is used to generate a structure~$H = (W,R,L)$
  as described next.
  Let~$W$ be the set of all states of~$\tl$.
  For every~$a \in \act$ and every~$s, t \in W$,
  let~$\prel{s}{R_a}{t}$ iff~$s$ contains a formula~$\pea{a}{\psi}$
  and there exists a path $x_0 = s, x_1, \dotsc, x_{k+1} = t$ in~$\tl$
  such that~$x_1$ is the (possibly virtual) successor of~$\pea{a}{\psi}$
  and each~$x_i, 1 \leq i \leq k$ is an $\talpha$- or a $\tbeta$-node.
  Thus state~$t$ is a ``saturation'' of~$x_1$ using only $\talpha$- and $\tbeta$-rules.
  Note that~$\prel{s}{R_a}{t}$ and~$\prel{s}{R_b}{t}$ is possible for~$a \not= b$,
  because two formulae~$\pea{a}{\psi} \in s$ and~$\pea{b}{\psi} \in s$ might have the same virtual successor:
  see point~\ref{enum_sev} of the \trea{}-rule.
  It is also possible that~$\prel{s}{R_a}{t}$ and~$\prel{s}{R_a}{u}$ for~$t \not= u$.

  If we consider the root~$r$ of~$\tl$ as a core-node for a moment,
  it is not hard to see that for every state~$s \in \tl$
  there exists a unique core-node~$x \in \tl$
  and a unique path~$\pi$ of the form $x_0 = x, x_1, \dotsc, x_k = s$ in~$\tl$
  such that either~$k = 0$ (and thus~$s = x$)
  or~$k > 0$ and each~$x_i, 0 \leq i \leq k-1$ is not a state.
  We set~$L(s)$ to be the union of all formulae of all nodes on~$\pi$.
  Intuitively, we form~$L(s)$ by adding back all the principal formulae
  of the $\talpha$- and $\tbeta$-rules
  which were applied to obtain~$s$ from~$x$.

  It is almost straightforward to check that~$H$ is a pre-Hintikka structure for~$\phi$.
  There are only two things that deserve extra comments:
  Firstly, it is not possible
  that~$\tl$ contains a \trert{}-node as it would be barred.
  Secondly, assume that~$y \in \tl$ is a \trar{}-node with principal formula~$\paa{\prp{\alpha}}{\varphi}$
  and~$s$ is a state such that~$y$ lies on the path~$\pi$ to~$s$
  that defines the set~$L(s)$, which contains~$\paa{\prp{\alpha}}{\varphi}$, as described above.
  Then either~$\varphi$ and~$\paa{\alpha}{\paa{\prp{\alpha}}{\varphi}}$ are contained in the child of~$y$ in~$\tl$,
  or -- as the first node~$x$ on~$\pi$ is a core-node with~$\tbbox_z = \emptyset$ --
  there exists another \trar{}-node on~$\pi$
  that also has~$\paa{\prp{\alpha}}{\varphi}$ as principal formula 
  and its child in~$\tl$ contains~$\varphi$ and~$\paa{\alpha}{\paa{\prp{\alpha}}{\varphi}}$.
  As the child of an $\talpha$-node that lies on~$\pi$ must lie on~$\pi$ too,
  in both cases, there is a node on~$\pi$ containing~$\varphi$ and~$\paa{\alpha}{\paa{\prp{\alpha}}{\varphi}}$.
  Thus~$\varphi$ and~$\paa{\alpha}{\paa{\prp{\alpha}}{\varphi}}$ are also contained in~$L(s)$.

  To show that~$H$ is even a Hintikka structure
  we use Lemma~\ref{lem_ex_fulfilling} to conclude~$H6$ as is shown next.

  Suppose~$\pea{\prp{\alpha}}{\varphi} \in L(s)$.
  If we also have~$\varphi \in L(s)$
  then~$(s, \pea{\prp{\alpha}}{\varphi}), (s, \varphi)$ is a fulfilling chain for~$(\varphi, \prp{\alpha}, s)$
  and we are done.
  Otherwise,
  the finiteness of the tableau
  and the fact that~$H$ is a pre-Hintikka structure
  give us a sequence~$\fchn = (s, \varphi_0), \dotsc, (s, \varphi_m)$
  such that:
  \begin{itemize}
  \item $\varphi_i \in \ppre{\pea{\prp{\alpha}}{\varphi}}$ and~$\varphi_i \in L(s)$ for all~$0 \leq i \leq m$
  \item $\varphi_0 = \pea{\prp{\alpha}}{\varphi}$ and $\varphi_m = \pea{a}{\varphi'}$ for some~$a \in \act$ and~$\varphi' \in \fml$
  \item $\varphi_i \pzz \varphi_{i+1}$ for all~$0 \leq i \leq m-1$.
  \end{itemize}
  Applying Lemma~\ref{lem_ex_fulfilling} for the state~$s$ and the formula~$\varphi_m = \pea{a}{\varphi'}$
  gives us a sequence~$\fchn' := (y_0, \psi_0), \dotsc, (y_n, \psi_n)$
  with the properties stated in Lemma~\ref{lem_ex_fulfilling}.
  Let~$y_n, \dotsc, y_{n+m}$ be an arbitrary path in~$\tl$ such that~$y_{n+m}$ is a state.
  Next, we replace each~$y_i, 1 \leq i \leq n$ in~$\fchn'$ with the first state~$s_i$
  that appears on the path $y_i, \dotsc, y_n, \dotsc, y_{n+m}$.

  It is easy to check that the combined sequence~$\fchn, \fchn'$
  is a fulfilling chain for~$(\varphi, \prp{\alpha}, s)$ in~$H$
  if we contract all consecutive repetitions of pairs.
  This concludes the proof.
\end{proof}

\begin{lemma}
  \label{lem_ex_fulfilling}
  Let~$y \in \tl$ be a node and~$\psi \in y$ a formula such that~$\psi \in \ppre{\pea{\prp{\alpha}}{\varphi}}$.
  There exists a finite sequence~$\fchn' = (y_0, \psi_0), \dotsc, (y_n, \psi_n)$ of pairs with~$n \geq 0$ such that:
  \begin{itemize}
  \item $y_0, \dotsc, y_n$ is a path in~$\tl$
  \item $y_i \in \tl$, $\psi_i \in \ppre{\varphi}$, and~$\psi_i \in y_i$ for all~$0 \leq i \leq n$
  \item $y_0 = y$, $\psi_0 = \psi$, $\psi_n = \varphi$, and~$\psi_i \neq \varphi$ for all~$0 \leq i \leq n-1$
  \item for all~$0 \leq i \leq n-1$, either $\psi_i = \psi_{i+1}$ or:
    if~$\psi_i = \pea{a}{\chi}$ for some~$a \in \act$ and~$\chi \in \fml$
    then~$y_i$ is a state else~$\psi_i \pzz \psi_{i+1}$.
  \end{itemize}
\end{lemma}
\begin{proof}
  We inductively construct~$\fchn'$ starting with~$(y_0, \psi_0) := (y, \psi)$.
  Most of the required properties of~$\fchn'$ follow directly from its construction
  and we leave it to the reader to check that they hold.
  \begin{step}
    \label{step_one}
    Let~$(y_i, \psi_i)$ be the last pair of~$\fchn'$.
    We distinguish three cases: either~$y_i$ is an $\talpha$- or $\tbeta$-node
    and~$\psi_i$ is not the principal formula in~$y_i$;
    or~$y_i$ is an $\talpha$- or $\tbeta$-node
    and~$\psi_i$ is the principal formula in~$y_i$;
    or~$y_i$ is a state.

    If~$y_i$ is an $\talpha$- or $\tbeta$-node
    and~$\psi_i$ is not the principal formula in~$y_i$,
    we set~$\psi_{i+1} := \psi_i$ and we choose~$y_{i+1}$ to be a successor of~$y_i$ in~$\tl$
    such that $\tuev_{y_i}(\psi_i, \pea{\prp{\alpha}}{\varphi}) = \tuev_{y_{i+1}}(\psi_{i+1}, \pea{\prp{\alpha}}{\varphi})$.
    Note that such a~$y_{i+1}$ always exists
    since the value of~$\tuev_{y_i}(\psi_i, \pea{\prp{\alpha}}{\varphi})$
    is determined by one of its open children during the construction of~$T$ and hence~$\tl$.
    But it does not have to be unique.
    We then repeat Step~\ref{step_one}.

    If~$y_i$ is an $\talpha$- or $\tbeta$-node
    and~$\psi_i$ is the principal formula in~$y_i$,
    we look at all pairs~$(x, \chi)$
    such that~$x$ is a child of~$y_i$ in~$\tl$
    and~$\psi_i$ is decomposed into~$\chi \in x$
    and~$\psi_i \pzz \chi$ holds.
    By construction of~$T$ and hence~$\tl$ there is at least one open child
    such that the corresponding pair~$(x, \chi)$
    obeys~$\tuev_{y_i}(\psi_i, \pea{\prp{\alpha}}{\varphi}) = \tuev_x(\chi, \pea{\prp{\alpha}}{\varphi})$.
    Let~$(y_{i+1}, \psi_{i+1})$ be such a pair.
    If~$\psi_{i+1} = \varphi$ we stop and return~$\fchn'$;
    otherwise we repeat Step~\ref{step_one}.

    If~$y_i$ is a state,
    it is not too hard to see that~$\psi_i$ must be of the form~$\pea{a}{\chi}$
    for some~$a \in \act$ and~$\chi \in \fml$.
    We set~$(y_{i+1}, \psi_{i+1}) := (x, \chi)$
    where~$x$ is the (possibly virtual) successor of~$\psi_i = \pea{a}{\chi}$
    and repeat Step~\ref{step_one}.
    Note that if~$x$ is a non-virtual successor of~$\psi_i$,
    we have~$\tuev_{y_i}(\psi_i, \pea{\prp{\alpha}}{\varphi}) = \tuev_{y_{i+1}}(\psi_{i+1}, \pea{\prp{\alpha}}{\varphi})$
    by construction of~$T$ and hence~$\tl$.
    Also note that if~$x$ is a virtual successor of~$\psi_i$
    then~$\psi_{i+1} = \chi$ is the core-formula of~$y_{i+1}$ by construction of~$T$ and hence~$\tl$.
  \end{step}
  The only way for Step~\ref{step_one} to terminate is
  by finding~$\psi_{i+1} = \varphi$.
  It is not difficult to see that
  the resulting (finite) sequence~$\fchn'$ fulfils all requirements and the proof is completed.
  Hence the rest of the proof shows that~$\fchn'$ as constructed by Step~\ref{step_one} is finite.
  Step~\ref{step_one} maintains the following invariant:
  \begin{description}
  \item[$(\dagger)$] For all appropriate~$i \in \Nat$
    we have~$\tuev_{y_i}(\psi_i, \pea{\prp{\alpha}}{\varphi}) = \tuev_{y_{i+1}}(\psi_{i+1}, \pea{\prp{\alpha}}{\varphi})$
    unless~$y_{i+1}$ is the \emph{virtual} successor of~$\psi_i \in y_i$.
  \end{description}
  In other words, the values of~$\tuev_{y_i}(\psi_i, \pea{\prp{\alpha}}{\varphi})$
  and~$\tuev_{y_{i+1}}(\psi_{i+1}, \pea{\prp{\alpha}}{\varphi})$ can differ only if~$(y_i, y_{i+1})$ is a backward edge in~$\tl$.
  We distinguish two cases: either~$\tuev_{y_0}(\psi_0, \pea{\prp{\alpha}}{\varphi})$ is undefined or it is defined.
  In both cases we show that the path~$y_0, y_1, \dotsc$
  can only have a finite number of backward edges.
  As every infinite path in~$\tl$ must use an infinite number of backward edges
  since~$T$ and~$\tm$ are finite trees,
  this proves that Step~\ref{step_one} terminates.

  \noindent{}\textbf{Case~1.} If~$\tuev_{y_0}(\psi_0, \pea{\prp{\alpha}}{\varphi})$ is undefined,
  the path~$y_0, y_1, \dotsc$ cannot contain a backward edge as shown next.
  Assume for a contradiction that~$y_i$ with~$i \geq 0$ is the first node
  such that~$(y_i, y_{i+1})$ is a backward edge.
  Since the initial $\tuev_{y_0}(\psi_0, \pea{\prp{\alpha}}{\varphi})$ was undefined,
  by~$(\dagger)$ we know that~$\tuev_{y_i}(\psi_i, \pea{\prp{\alpha}}{\varphi})$ is undefined.
  But~$y_i$ is a state and as~$\psi_i \in y_i$,
  which must be of the form~$\pea{a}{\chi}$ for some~$a \in \act$ and~$\chi \in \fml$,
  has a virtual successor~$z$,
  $\tuev_{y_i}(\psi_i, \pea{\prp{\alpha}}{\varphi})$ is defined
  to be the height of~$z$ by the application of the \trea{}-rule to~$y_i$
  during the construction of the tableau.
  Thus~$\tuev_{y_i}(\psi_i, \pea{\prp{\alpha}}{\varphi})$ is both defined and undefined,
  which is a contradiction.

  \noindent{}\textbf{Case~2.} If~$h := \tuev_{y_0}(\psi_0, \pea{\prp{\alpha}}{\varphi})$ is defined,
  the path~$y_0, y_1, \dotsc$ can only contain a finite number of backward edges as shown next.
  Let~$y_i$ with~$i \geq 0$ be the first node
  such that~$(y_i, y_{i+1})$ is a backward edge.
  If no such node exists, we are obviously done.
  Otherwise, we have~$\tuev_{y_i}(\psi_i, \pea{\prp{\alpha}}{\varphi}) = h$ by~$(\dagger)$.
  This means by construction of the tableau
  that there exists a set~$\Delta \subseteq \fml$ such that~$(\psi_{i+1}, \{ \psi_{i+1} \} \cup \Delta) = \thc_{y_i}[h]$.
  Thus~$y_{i+1}$ is the $h^{\mathrm{th}}$ core-node (child of a \trea{}-node)
  on the path from the root~$r$ to~$y_i$ in~$\tl$
  and we have~$\llen(\thc_{y_{i+1}}) = h$ by construction of~$\thc$.

  If~$\tuev_{y_{i+1}}(\psi_{i+1}, \pea{\prp{\alpha}}{\varphi})$ had a value equal to or greater than~$h$
  then the \trea{}-rule would cause the parent of~$y_{i+1}$ in~$\tl$
  to be marked as closed
  since~$\psi_{i+1}$ is the core-formula of~$y_{i+1}$;
  but we know this is not the case.
  Hence~$\tuev_{y_{i+1}}(\psi_{i+1}, \pea{\prp{\alpha}}{\varphi})$ is either undefined or
  has a value~$h'$ that is strictly smaller than~$h$.

  If~$\tuev_{y_{i+1}}(\psi_{i+1}, \pea{\prp{\alpha}}{\varphi})$ is undefined,
  we can prove exactly as in Case~1 that the path~$y_{i+1}, y_{i+2}, \dotsc$ cannot contain a backward edge.
  On the other hand,
  if~$h' := \tuev_{y_{i+1}}(\psi_{i+1}, \pea{\prp{\alpha}}{\varphi})$ is defined,
  we can inductively repeat the arguments in Case~2
  for the sequence $(y_{i+1}, \psi_{i+1}), (y_{i+2}, \psi_{i+2}), \dotsc$.
  The induction is well-defined because of~$h' < h$,
  meaning that eventually this inductive argument must terminate
  because all such $h$-values must be in~$\Natp$.
\end{proof}

\paragraph{\bf Completeness}

\begin{definition}
  \label{def_chain}
  Let~$M = (W,R,V)$ be a model, $w \in W$ a state
  and~$\varphi \in \fml$ a formula of the form~$\varphi = \pea{\alpha_1}{\dotsc \pea{\alpha_k}{\psi}}$
  for some~$k > 0$ and $\alpha_1, \dots, \alpha_k \in \prg$ and~$\psi \in \fml$.
  A \emph{witness chain for~$(\varphi, \psi, M, w)$}
  is a finite sequence~$(w_0, \psi_0), \dotsc, (w_n, \psi_n)$
  of world-formula pairs with~$n > 0$ such that:
  \begin{enumerate}
  \item $w_i \in W$, $\psi_i \in \ppre{\psi}$, and~$\psr{M,w_i}{\psi_i}$ for all~$0 \leq i \leq n$
  \item $w_0 = w$, $\psi_0 = \varphi$, $\psi_n = \psi$,
    and~$\psi_i \neq \psi$ for all~$0 \leq i \leq n-1$
  \item $\forall i, j \in \{ 0, \dotsc, n \}.\: i \not= j \ximp (w_i, \psi_i) \not= (w_{i+1}, \psi_{i+1})$
  \item for all~$0 \leq i \leq n-1$,
    if~$\psi_i = \pea{a}{\chi}$ for some~$a \in \act$ and~$\chi \in \fml$
    then~$\psi_{i+1} = \chi$ and~$\prel{w_i}{R_a}{w_{i+1}}$;
    otherwise~$\psi_i \pzz \psi_{i+1}$ and~$w_i = w_{i+1}$.
  \end{enumerate}
\end{definition}
\begin{prop}
  \label{prop_lastelem}
  In the setting of Def.~\ref{def_chain}, we have:
  \begin{enumerate}
  \item for every~$1 \leq i \leq k$ there exists an~$m < n$
    such that~$(w_0, \psi_0), \dotsc, (w_m, \psi_m)$
    is a witness chain for~$(\varphi, \pea{\alpha_i}{\dotsc \pea{\alpha_k}{\psi}}, M, w)$
  \item if~$\alpha_k = \prp{\beta}$ for some~$\beta \in \prg$ then~$\psi_{n-1} = \pea{\prp{\beta}}{\psi}$.
  \end{enumerate}
\end{prop}
\begin{proposition}
  \label{prop_chain}
  Let~$M = (W,R,V)$ be a model, $w \in W$ a state
  and~$\varphi \in \fml$ a formula of the form~$\varphi = \pea{\alpha_1}{\dotsc \pea{\alpha_k}{\psi}}$
  for some~$k > 0$ and $\alpha_1, \dots, \alpha_k \in \prg$ and~$\psi \in \fml$.
  If~$\psr{M,w}{\varphi}$ then there exists a witness chain for~$(\varphi, \psi, M, w)$.
\end{proposition}

From now on,
let~$\Gamma_y$ denote the set of formulae of a node~$y \in T$.
We say that a finite set of formulae~$\Gamma$ is satisfiable
iff~$\bigwedge_{\varphi \in \Gamma}\varphi$ is satisfiable.
\begin{lemma}
  \label{lemma_focus}
  Let~$x \in T$ with~$\tbdia_x = \emptyset$ and principal formula~$\varphi \in \fmlean$
  of the form~$\varphi = \pea{\alpha_1}{\dotsc \pea{\alpha_k}{\psi}}$
  for some~$k > 0$ and $\alpha_1, \dots, \alpha_k \in \prg$ and~$\psi \in \fml \setminus \fmlea$.
  Let~$M = (W,R,V)$ be a model and~$w \in W$ a world such that~$(M, w)$ satisfies~$\Gamma_x$.
  Furthermore let~$\fchn = (w_0, \psi_0), \dotsc, (w_n, \psi_n)$ be a witness chain for~$(\varphi, \psi, M, w)$.
  Then there exists a finite path~$\pi = z_0, z_1, \dotsc, z_m$ in~$T$
  with the following properties:
  \renewcommand{\theenumi}{(\roman{enumi})}
  \renewcommand{\labelenumi}{\theenumi}
  \begin{enumerate}
  \item $m \leq n$, $z_0 = x$, $\tbdia_{z_m} = \emptyset$, and the only state (if any) is~$z_m$
  \item $w_i = w$, $\psi_i \in z_i$, and~$(M, w)$ satisfies~$\Gamma_{z_i}$ for all~$0 \leq i \leq m$
  \item $\psi_i \in \fmlean$ is the principal formula of~$z_i$ for all~$0 \leq i \leq m-1$
  \item $\psi_m = \psi$ or~$\psi_m = \pea{a}{\chi}$ for some~$a \in \act$ and~$\chi \in \fml$.
  \end{enumerate}
\end{lemma}
\begin{proof}
  We inductively construct~$\pi$ starting with~$z_0 = x$,
  such that the following invariant holds:
  \begin{description}
  \item[$(\sharp)$] $m < n$ and for all~$0 \leq i \leq m$:
    $w_i = w$ and~$(M, w)$ satisfies~$\Gamma_{z_i}$ and~$\psi_i \in \fmlean$ is the principal formula of~$z_i$.
  \end{description}
  Note that~$(\sharp)$ holds for the initial path~$\pi = z_0$.
  Also note that if~$\pi$ fulfils~$(\sharp)$
  then no node on~$\pi$ can be a state and
  and $\psi_i \in z_i$ for all~$0 \leq i \leq m$.
  \begin{step}
    \label{step_two}
    Let~$z_m$ be the last node of~$\pi$.
    It cannot be an \trid{}-node because it is satisfiable,
    nor a \trert{}-node for the following reason:
    Assume that~$z_m$ were a \trert{}-node.
    Then~$\psi_m \in \tbdia_{z_m}$ due to the \trert{}-rule
    and there must be an ancestor node~$z$ of~$z_m$ in~$T$
    which inserted~$\psi_m$ into the~$\tbdia$ of its child
    such that~$\psi_m$ is contained in the~$\tbdia$ of all nodes
    between~$z$ (exclusive) and~$z_m$ (inclusive).
    As~$\tbdia_{z_0} = \emptyset$ by assumption,
    the node~$z$ must lie on~$\pi$,
    \ie{}~$z = z_{m'}$ for some~$m' < m$.
    Due to the tableau rules and the fact that~$z$ inserted~$\psi_m$, 
    the node~$z$ must be a \trero{}-node with principal formula~$\psi_m$;
    but that -- together with~$(\sharp)$ --
    entails~$(w_{m'},\psi_{m'}) = (w, \psi_m) = (w_m, \psi_m)$
    which is not possible because~$\fchn$ is a witness chain.
    Hence~$z_m$ is a not a \trert{}-node.

    Let~$z_{m+1}$ be the child of~$z_m$
    where~$\psi_m$ is decomposed into~$\psi_{m+1}$.
    Such a child must exist because we have~$m < n$ and~$\psi_m \pzz \psi_{m+1}$
    due to the definition of the witness chain~$\fchn$ and the fact that~$\psi_m \in \fmlean$.
    The same reasoning also gives us~$w = w_m = w_{m+1}$ and~$\psr{M,w}{\psi_{m+1}}$.
    Moreover, the set~$\Gamma_{z_m}$ is satisfied by~$(M,w)$ by~$(\sharp)$
    and~$\Gamma_{z_{m+1}} = (\Gamma_{z_m} \setminus \{ \psi_m \}) \cup \{ \psi_{m+1} \}$ by construction of the tableau~$T$.
    Hence the set~$\Gamma_{z_{m+1}}$ is satisfied by~$(M,w)$.

    Now we distinguish whether or not~$\psi_{m+1}$ is a \fean{}-formula.

    If~$\psi_{m+1}$ is a \fean{}-formula,
    it must be the principal formula of~$z_{m+1}$
    due to the tableau rules and the fact that we have~$\psi_m \in \fmlean$.
    Moreover, we have~$m+1 < n$ because~$\psi_{m+1} \not= \psi = \psi_n$ and~$\psi \notin \fmlea$.
    Thus our invariant~$(\sharp)$ for~$\pi$ extended by~$\psi_{m+1}$ still holds
    and we repeat Step~\ref{step_two}.

    If~$\psi_{m+1}$ is not a \fean{}-formula,
    we have~$\tbdia_{z_{m+1}} = \emptyset$
    due to the tableau rules and the fact that~$\psi_m \in \fmlean$.
    Furthermore, we have~$\psi_m = \psi$ or~$\psi_m = \pea{a}{\chi}$ for some~$a \in \act$ and~$\chi \in \fml$
    because~$\fchn$ is a witness chain.
    Thus~$\pi$ extended by~$\psi_{m+1}$ fulfils all the required properties of the lemma
    which concludes the proof in this case.
  \end{step}
  As~$\fchn$ is finite,
  Step~\ref{step_two} must terminate after a finite number of repetitions
  which means that we have found a path~$\pi$ that proves this lemma.
\end{proof}

\begin{lemma}
  \label{lemma_findstate}
  Let~$x \in T$ with~$\tbdia_x = \emptyset$
  and~$M = (W,R,V)$ be a model and~$w \in W$ a world
  such that~$(M, w)$ satisfies~$\Gamma_x$.
  Then there exists a finite path~$\pi = z_0, z_1, \dotsc, z_n$ in~$T$
  with the following properties:
  $z_0 = x$, $z_n$ is the only state on~$\pi$, and~$(M, w)$ satisfies~$\Gamma_{z_i}$ for all~$0 \leq i \leq n$.
\end{lemma}
\begin{proof}
  We inductively construct~$\pi$ starting with~$z_0 = x$
  such that the following invariant holds:
  \begin{description}
  \item[$(\ddagger)$] $(M, w)$ satisfies~$\Gamma_y$ for every node~$y$ on~$\pi$
    and the last node~$z_i$ of~$\pi$ has~$\tbdia_{z_i} = \emptyset$.
  \end{description}
  Note that the initial~$\pi = z_0$ fulfils the invariant by assumption.
  \begin{step}
    \label{step_three}
    Let~$z_i$ be the last node of~$\pi$.
    If~$z_i$ is a state, we stop and return~$\pi$.
    Otherwise, we distinguish two cases:
    either the principal formula of~$z_i$ is not a \fea{}-formula
    or it is a \fea{}-formula.

    If the principal formula of~$z_i$ is not a \fea{}-formula,
    we choose~$z_{i+1}$ to be a successor of~$z_i$ in~$T$
    such that~$(M, w)$ satisfies~$\Gamma_{z_{i+1}}$.
    The existence of~$z_{i+1}$ is guaranteed by Prop.~\ref{prop_axioms},
    the fact that~$(M, w)$ satisfies~$\Gamma_{z_i}$ by~$(\ddagger)$,
    and the fact that~$z_i$ cannot be an \trid{}-node because~$z_i$ is satisfiable
    nor a \trert{}-node because~$z_i$'s principal formula is not a \fea{}-formula.
    As~$z_i$'s principal formula is not a \fea{}-formula and~$\tbdia_{z_i} = \emptyset$ by~$(\ddagger)$, 
    we also have~$\tbdia_{z_{i+1}} = \emptyset$ by a simple inspection of the tableau rules.
    We then repeat Step~\ref{step_three}.
    
    If the principal formula~$\varphi$ of~$z_i$ is a \fea{}-formula,
    it is also a \fean{}-formula because~$z_i$ is not a state.
    Hence it must be of the form~$\varphi = \pea{\alpha_1}{\dotsc \pea{\alpha_k}{\psi}}$
    for some~$k > 0$ and~$\alpha_1, \dots, \alpha_k \in \prg$ and~$\psi \in \fml \setminus \fmlea$.
    As~$(M, w)$ satisfies~$\Gamma_{z_i}$ by~$(\ddagger)$ and~$\varphi \in \Gamma_{z_i}$,
    we have~$\psr{M,w}{\varphi}$.
    Thus Prop.~\ref{prop_chain} gives us a sequence~$\fchn := (w_0, \psi_0), \dotsc, (w_n, \psi_n)$ 
    with the properties stated in Prop.~\ref{prop_chain}.

    Next we apply Lemma~\ref{lemma_focus} to~$z_i$
    and obtain a path~$\tau$ with the properties of Lemma~\ref{lemma_focus}.
    Finally,
    the new~$\pi$ is obtained from the old~$\pi$ by appending~$\tau$
    -- minus the first node~$z_i$
    which is already the last node of~$\pi$ -- to the old~$\pi$.
    As $(M, w)$ satisfies~$\Gamma_y$ for all~$y$ on~$\tau$
    and the last node~$y'$ on~$\tau$ has~$\tbdia_{y'} = \emptyset$,
    the new~$\pi$ fulfils~$(\ddagger)$.
    We then repeat Step~\ref{step_three}.
  \end{step}
  As~$T$ is finite, it is easy to see that Step~\ref{step_three} terminates,
  meaning that the last node~$z_n$ of the finite path~$\pi$ is the only state on~$\pi$.
\end{proof}

\begin{lemma}
  \label{lemma_marked}
  For every closed node~$x = \tnode{\Gamma}{\dotsb}{\dotsb}$ in~$T$,
  the set~$\Gamma_x$ is not satisfiable.
  In particular, if~$r$ is closed then~$\phi$ is not satisfiable.
\end{lemma}
\begin{proof}
  We use well-founded induction on the (strict) descendant relation of~$T$.
  As~$T$ is a finite tree, the descendant relation is clearly well-founded.
  Thus we can use the following induction hypothesis for every node~$x \in T$:
  \begin{description}
  \item[IH:] for every descendant~$y$ of~$x$,
    if~$y$ is closed then the set~$\Gamma_y$ is not satisfiable.
  \end{description}

  If a leaf~$x \in T$ is closed,
  it must be an \trid{}-node
  as a state with no children is always open.
  Hence, our theorem follows from the fact
  that~$\{ p, \pnot{p} \} \subseteq x$ for some~$p \in \proptn$.
  Note that this can be seen as the base case of the induction
  as leaves do not have descendants.

  If~$x$ is a closed $\talpha$-node
  then its child must be closed as well
  so we can apply the induction hypothesis
  and the claim follows from the fact
  that -- in the sense of Table~\ref{tab_alphabeta} --
  the formulae of the form~$\peq{\talpha}{\pand{\talpha_1}{\talpha_2}}$ are valid
  (Prop.~\ref{prop_axioms}).

  If~$x$ is a closed $\tbeta$-node
  then both children are closed as well
  so we can apply the induction hypothesis
  and the claim follows from the fact
  that -- in the sense of Table~\ref{tab_alphabeta} --
  the formulae of the form~$\peq{\tbeta}{\por{\tbeta_1}{\tbeta_2}}$ are valid
  (Prop.~\ref{prop_axioms}).
  Note that~$x$ cannot be a \trert{}-node as it would not be closed in this case.

  If~$x$ is a closed \trea{}-node
  (\ie{} a closed state)
  then it has at least one child and there are three possibilities
  for why it was marked as closed by the \trea{}-rule:
  \renewcommand{\theenumi}{(\arabic{enumi})}
  \renewcommand{\labelenumi}{\theenumi}
  \begin{enumerate}
  \item Some child~$x_0$ of~$x$ is closed.
  \item Some child~$x_0$ of~$x$ is barred.
  \item Some open child~$x_0$ of~$x$ with core-formula~$\varphi$
    has $\tuev_{x_0}(\varphi, \pea{\prp{\alpha}}{\psi}) > h := \llen(\thc_x)$
    for some~$\alpha \in \prg$ and~$\psi \in \fml$ with~$\varphi \in \ppre{\pea{\prp{\alpha}}{\psi}}$.
  \end{enumerate}

  \noindent{}\textbf{Case~1.}
  In the first case,
  it is not too hard to see
  that the satisfiability of~$\Gamma_x$ implies the satisfiability of~$\Gamma_{x_0}$
  since the \trea{}-rule preserves satisfiability from parent to child.
  By the induction hypothesis,
  we know that~$\Gamma_{x_0}$ is not satisfiable,
  therefore~$\Gamma_x$ cannot be satisfiable either.

  \noindent{}\textbf{Case~2.}
  In the second case,
  we assume that~$\Gamma_{x_0}$ is satisfiable and derive a contradiction.
  We can then prove the claim as in the first case.

  So, for a contradiction,
  let~$M = (W,R,V)$ be a model and~$w \in W$ a world
  such that~$(M, w)$ satisfies~$\Gamma_{x_0}$.
  As~$\tbdia_{x_0} = \emptyset$ by the \trea{}-rule,
  we can apply Lemma~\ref{lemma_findstate}
  which gives us a path~$\pi$ in~$T$ with the properties stated in Lemma~\ref{lemma_findstate}.
  Let~$y$ be the last node of~$\pi$, hence~$y$ is a state.
  It is a descendant of~$x_0$,
  therefore the induction hypothesis applies to it.
  By Lemma~\ref{lemma_findstate}, $(M, w)$ satisfies~$\Gamma_y$,
  hence~$y$ cannot be closed;
  but this means that~$y$ must be open 
  as states can only be closed or open by the \trea{}-rule.
  It is now easy to see that all nodes on~$\pi$ must also be open 
  due to the construction of the variable~$\tmrk$ in the $\talpha$- and $\tbeta$-rules.
  But this is a contradiction to the assumption
  that~$x_0$, which is the first node on~$\pi$, is barred.

  \noindent{}\textbf{Case~3.}
  In the third case,
  we assume that~$\Gamma_{x_0}$ is satisfiable and derive a contradiction.
  We can then prove the claim as in the first case.

  So, for a contradiction,
  let~$M = (W,R,V)$ be a model and~$w \in W$ a world
  such that~$(M, w)$ satisfies~$\Gamma_{x_0}$.
  In particular, we have~$\psr{M,w}{\varphi}$ by assumption since~$\varphi \in x_0$.
  As~$\varphi \in \ppre{\pea{\prp{\alpha}}{\psi}}$,
  it is of the form~$\varphi = \pea{\alpha_1}{\dotsc \pea{\alpha_{k-1}}{\pea{\prp{\alpha}}{\psi}}}$
  for some~$\alpha_1, \dots, \alpha_{k-1} \in \prg$.
  Furthermore,
  let~$\psi$ be of the form~$\psi = \pea{\alpha_{k+1}}{\dotsc \pea{\alpha_{k+l}}{\psi'}}$
  for some~$\alpha_{k+1}, \dots, \alpha_{k+l} \in \prg$ and~$\psi' \in \fml \setminus \fmlea$.
  Note that~$l = 0$ is possible:
  in this case we already have~$\psi \in \fml \setminus \fmlea$.

  Applying Prop.~\ref{prop_chain} to~$M$ and~$\varphi = \pea{\alpha_1}{\dotsc \pea{\alpha_{k+l}}{\psi'}}$
  with~$\alpha_k := \prp{\alpha}$
  gives us a witness chain~$\fchn = (w_0, \psi_0), \dotsc, (w_n, \psi_n)$ for~$(\varphi, \psi', M, w)$.
  According to Prop~\ref{prop_lastelem},
  there exists an~$n' \leq n$ with~$\psi_{n'} = \psi = \pea{\alpha_{k+1}}{\dotsc \pea{\alpha_{k+l}}{\psi'}}$
  and~$\psi_{n'-1} = \pea{\prp{\alpha}}{\psi}$.
  Our plan is to ``walk down'' the tableau~$T$ -- starting from~$x_0$ --
  in a way that is ``consistent'' with~$\fchn$
  which will lead to a contradiction when we ``reach''~$\psi_{n'}$.

  As~$\tbdia_{x_0} = \emptyset$ by the \trea{}-rule,
  we can apply Lemma~\ref{lemma_focus}
  which gives as a path~$\pi_1 = z_0, z_1, \dotsc, z_m$ in~$T$
  with the properties stated in Lemma~\ref{lemma_focus}.
  We can then apply Lemma~\ref{lemma_findstate} to~$z_m$
  which gives us a path~$\pi_2$ with the properties stated in Lemma~\ref{lemma_findstate}.
  Let~$s$ be the last node of~$\pi_2$, hence~$s$ is a state.
  It is a descendant of~$x_0$,
  therefore the induction hypothesis applies to it.
  Thus~$s$ cannot be closed because~$(M, w)$ satisfies~$\Gamma_s$ by Lemma~\ref{lemma_findstate};
  but this means that~$s$ must be open 
  as states can only be closed or open by the \trea{}-rule.
  If we join~$\pi_1$ and~$\pi_2$ to obtain~$\pi$,
  it is now easy to see that all nodes on~$\pi$ must also be open 
  due to the construction of the variable~$\tmrk$ in the $\talpha$- and $\tbeta$-rules.

  By assumption we have~$\tuev_{x_0}(\varphi, \pea{\prp{\alpha}}{\psi}) > h$.
  As all nodes on~$\pi_1$ are open
  and~$\psi_i \in \fmlean$ is the principal formula of~$z_i$ for all~$0 \leq i \leq m-1$,
  we also have~$\tuev_{z_i}(\psi_i, \pea{\prp{\alpha}}{\psi}) > h$ for all~$0 \leq i \leq m-1$
  by definition of the $\talpha$- and $\tbeta$-rules.
  We now distinguish whether or not~$n' \leq m$.

  If~$n' \leq m$ then we have $\tuev_{z_{n'-1}}(\pea{\prp{\alpha}}{\psi}, \pea{\prp{\alpha}}{\psi}) > h$
  as~$\psi_{n'-1} = \pea{\prp{\alpha}}{\psi}$;
  but as~$\pea{\prp{\alpha}}{\psi}$ is the principal formula of~$z_{n'-1}$,
  this is only possible if the first child of~$z_{n'-1}$,
  which is~$z_{n'}$ as~$\psi_{n'} = \psi$ by definition of~$\psi_{n'}$, is not open
  according to the construction of~$\tuev$ in the \trero{}-rule.
  This, however, is a contradiction to the fact that all nodes on~$\pi_1$,
  in particular~$z_{n'}$, are open.

  If~$n' > m$, we must have~$\psi_m = \pea{a}{\chi}$ for some~$a \in \act$ and~$\chi \in \fml$
  as~$\psi_m = \psi'$ is clearly not possible.
  Furthermore, we have~$\tuev_{z_m}(\pea{a}{\chi}, \pea{\prp{\alpha}}{\psi}) > h$
  by definition of the $\talpha$- and $\tbeta$-rules.
  As~$\psi_m$ is the first node on~$\pi_2$ 
  and all nodes on~$\pi_2$ are open,
  we also have~$\tuev_y(\pea{a}{\chi}, \pea{\prp{\alpha}}{\psi}) > h$ for all nodes~$y$ on~$\pi_2$
  by definition of the $\talpha$- and $\tbeta$-rules.
  In particular, we have~$\tuev_s(\pea{a}{\chi}, \pea{\prp{\alpha}}{\psi}) > h$.
  Let~$x_1$ be the (possibly virtual) successor of~$\pea{a}{\chi} \in s$,
  that contains~$\psi_{m+1} = \chi$.
  Then~$\fchn' := (w_{m+1}, \psi_{m+1}), \dotsc, (w_n, \psi_n)$ is clearly 
  a witness chain for~$(\psi_{m+1}, \psi', M, w_{m+1})$
  which is strictly shorter than~$\fchn$
  and still contains~$\psi_{n'}$ and~$\psi_{n'-1}$.
  Note that~$n' > m+1$ as~$\psi_m = \pea{a}{\chi} \neq \pea{\prp{\alpha}}{\psi}$.
  Additionally, we make the following two claims:
  \begin{enumerate}
  \item $\tuev_{x_1}(\psi_{m+1}, \pea{\prp{\alpha}}{\psi}) > h$ 
    and~$x_1$ is a descendant of~$x_0$
    (\ie{} the induction hypothesis holds in the subtree rooted at~$x_1$).
  \item $(M, w_{m+1})$ satisfies~$\Gamma_{x_1}$.
  \end{enumerate}
  Before we prove the two claims, we show their consequences:
  Basically, the two claims and the properties of~$\fchn'$ allow us
  to inductively repeat the proof
  for~$x_1$, $w_{m+1}$, $\psi_{m+1}$, and~$\fchn'$ instead of~$x_0$, $w$, $\varphi$, and~$\fchn$, respectively.
  As~$\fchn'$ is strictly shorter than~$\fchn$,
  this is possible only a finite number of times.
  Hence we must eventually end up in the case ``$n' \leq m$'' of the proof
  that yields a contradiction.
  Therefore the only thing left is to show that the two claims hold.

  \noindent{}\textbf{Claim~1.}
  We distinguish whether~$x_1$ is a virtual successor of~$\pea{a}{\chi} \in s$ or not.
  
  If~$x_1$ is not virtual,
  that is it is a child of~$s$ in~$T$,
  it is obviously a descendant of~$x_0$
  as every node -- in particular~$s$ -- on~$\pi$ is a descendant of~$x_0$.
  Furthermore,
  it follows directly from~$\tuev_s(\pea{a}{\chi}, \pea{\prp{\alpha}}{\psi}) > h$ and~$\psi_{m+1} = \chi$
  and the definition of the \trea{}-rule
  that~$\tuev_{x_1}(\psi_{m+1}, \pea{\prp{\alpha}}{\psi}) > h$.

  If~$x_1$ is a virtual successor,
  a glance at the definition of~$\tuev_s$ in the \trea{}-rule reveals
  that~$x_1$ must lie on the path from~$x_0$ to~$s$ (it could be~$x_0$)
  as we have $\tuev_s(\pea{a}{\chi}, \pea{\prp{\alpha}}{\psi}) > h$ and $h = \llen(\thc_x)$.
  Thus~$x_1$ is a descendant of~$x$
  and has~$\tuev_{x_0}(\chi, \pea{\prp{\alpha}}{\psi}) > h$
  as we have already established this on our way from~$x_0$ down to~$s$.
  
  \noindent{}\textbf{Claim~2.}
  By definition of the \trea{}-rule,
  $\Gamma_{x_1}$ is of the form~$\psi_{m+1} \cup \Delta$ where~$\paa{a}{\Delta} \subseteq \Gamma_s$.
  We know~$\psr{M,w_{m+1}}{\psi_{m+1}}$ because of the properties of~$\fchn$.
  We also know that~$(M,w_m)$ in particular satisfies~$\paa{a}{\Delta}$
  since we have established that~$\Gamma_s \supseteq \paa{a}{\Delta}$
  is satisfied by~$(M,w)$ and~$w = w_m$.
  As~$w_{m+1}$ is a successor world of~$w$ (\ie{}~$\prel{w}{R_a}{w_{m+1}}$),
  this implies that~$(M,w_{m+1})$ satisfies~$\Delta$,
  and hence~$\Gamma_{x_1}$.
\end{proof}

\noindent{}\textbf{Theorem~\ref{theo_completeness}\ }
\emph{If the root~$r \in T$ is not open then~$\phi$ is not satisfiable.}
\begin{proof}
  If~$r$ is closed, the claim follows directly from Lemma~\ref{lemma_marked}.
  If~$r$ is barred, 
  we assume that~$\Gamma_{x_0}$ is satisfiable and derive a contradiction.

  So, for a contradiction,
  let~$M = (W,R,V)$ be a model and~$w \in W$ a world
  such that~$(M, w)$ satisfies~$\Gamma_r = \phi$.
  As~$\tbdia_r = \emptyset$ by construction of~$T$,
  we can apply Lemma~\ref{lemma_findstate}
  which gives us a path~$\pi$ with the properties stated in Lemma~\ref{lemma_findstate}.
  Let~$y$ be the last node of~$\pi$, hence~$y$ is a state.
  It cannot be closed because of Lemma~\ref{lemma_marked}
  and the fact that~$(M, w)$ satisfies~$\Gamma_y$;
  but this means that~$y$ must be open 
  as states can only be closed or open by construction.
  It is easy to see that all nodes on~$\pi$ must also be open 
  due to the construction of the variable~$\tmrk$ in the $\talpha$- and $\tbeta$-rules.
  But this is a contradiction to the assumption
  that~$r$, which is the first node on~$\pi$, is barred.
\end{proof}

\end{document}